\documentclass[usletter,10spt]{article}

\usepackage[english]{babel}
\usepackage{amssymb}
\usepackage{amsmath}
\usepackage[dvips]{graphicx}
\usepackage{amsthm}
\usepackage[T1]{fontenc}
\usepackage{ae,aecompl}
\usepackage{subfigure}
\usepackage[round]{natbib}
\usepackage[colorlinks]{hyperref}
\usepackage{appendix}
\hypersetup{colorlinks,breaklinks=true,citecolor=blue,filecolor=blue,linkcolor=blue,urlcolor=blue,pdfauthor={C.~Guervilly}}

\newcommand{\vect}[1]{\mathbf{#1}}
\newcommand{\pdt}[1]{\frac{\partial #1}{\partial t}}

\newcommand{\pleft}{\left(}
\newcommand{\pright}{\right)}
\newcommand{\Rey}{\textrm{Re}}
\newcommand{\Rm}{\textrm{Rm}}
\newcommand{\Pm}{\textrm{Pm}}

\begin{document}

\title{Self-consistent simulations of a von K\'arm\'an type dynamo in a 
spherical domain with metallic walls}

\author{C\'eline Guervilly \& Nicholas H. Brummell \vspace{0.2cm}
 \\ {\small Department of Applied Mathematics and Statistics, Baskin School of Engineering,}
 \\ {\small University of California, Santa Cruz, CA95064, USA}}

\maketitle

\begin{abstract}
We have performed numerical simulations of 
boundary-driven dynamos using a three-dimensional non-linear magnetohydrodynamical 
model in a spherical shell geometry. A conducting fluid of magnetic Prandtl number $\Pm=0.01$ 
is driven into motion by the counter-rotation of the two hemispheric walls.
The resulting flow is of von K\'arm\'an type, consisting of
a layer of zonal velocity close to the outer wall and 
a secondary meridional circulation.
Above a certain forcing threshold, the mean flow is unstable to non-axisymmetric 
motions within an equatorial belt.
For fixed forcing above this threshold,
we have studied the dynamo properties of this flow. 
The presence of a conducting outer wall is essential to the existence 
of a dynamo at these parameters.
We have therefore studied
the effect of changing the material parameters of the wall
(magnetic permeability, electrical conductivity, and thickness) on the dynamo.
In common with previous studies, we find that
dynamos are obtained only when either the conductivity or the permeability
is sufficiently large.
However, we find that the effect of these two parameters on the dynamo process are different and can even compete to the detriment of the dynamo.
Our self-consistent approach allow us to analyze 
in detail the dynamo feedback loop.
The dynamos we obtain are typically dominated by an axisymmetric toroidal magnetic field
and an axial dipole component.
We show that the ability of the outer shear layer to produce a strong
toroidal field depends critically on the presence of a conducting outer wall, which shields 
the fluid from the vacuum outside.  
The generation of the axisymmetric poloidal field, on the other hand,
occurs in the equatorial belt and does not depend on the wall properties.
\end{abstract}

\section{Introduction}
An electrically conducting fluid driven by viscous forcing exerted at a boundary
generates a dynamo if the fluid's magnetic properties -- electrical conductivity and 
magnetic permeability -- and flow properties can amplify an initial weak magnetic field
and ultimately sustain a magnetic field of significant amplitude.
This is the most efficient type of forcing to convert the power
applied to the system into kinetic energy available for the dynamo, and so is 
preferred in laboratory experiments designed to study dynamo action, such as liquid metal experiments. 
In most of these experiments the energy injection scale is the largest scale of the system.

Recently, results from a boundary-driven dynamo experiment,
the von K\'arm\'an Sodium (VKS) experiment located in Cadarache, France,
have shown that the magnetic properties of the boundaries also greatly affect the ability of 
the flow to maintain a dynamo \citep{Mon07}.
The VKS experiment consists of a cylindrical container filled with liquid sodium, with two counter-rotating impellers
at either end. The mechanical forcing exerted by the impellers on the liquid sodium
drives a highly turbulent flow. 
For a sufficiently strong mechanical forcing, dynamo action has been observed that sustains 
a large-scale magnetic field despite the unconstrained and turbulent nature of the flow.
Furthermore, the axisymmetry of the sustained magnetic field (an axial dipole) implies that
the turbulent motions are involved in the dynamo process \citep[\emph{e.g.}][]{Pet07}.
This is an important result in the study of natural dynamos, which 
operate at very large Reynolds numbers, and mostly produce large-scale
magnetic fields.
However, dynamo action is only observed 
in the VKS experiment when the impellers are made of soft iron --
a material with high magnetic permeability, which produces a discontinuity 
in the magnetic field between the fluid and the impellers.
At the highest achievable mechanical forcing in the experiment, dynamo action has never 
been observed with either stainless steel or copper impellers \citep{Aum08}.
Consequently, elucidating the effect of changes of magnetic permeability of the impellers on the dynamo, and more 
generally of magnetic boundary conditions,
is critical to understanding how the dynamo mechanism
operates in the VKS dynamo experiment. 
This problem is also crucial in other shear-driven systems, such as
the plasma Couette experiment in Madison, Wisconsin \citep{Spe09}, and the spherical Couette liquid sodium experiment 
in College Park, Maryland \citep{Zim11}. 

The effect of magnetic boundary conditions on dynamo action has been investigated in numerical simulations
for both von K\'arm\'an type flow (between coaxial rotating disks) and Ponomarenko type flow 
(cylindrical helical flow) \citep{Pono73}.
Unfortunately, computational limitations prevent numerical simulations from reproducing 
the same level of turbulence obtained in laboratory experiments.
Numerical models include the large-scale mean velocity component, and
sometimes smaller scales with  typical viscous length scales much larger than natural or experimental dynamos
due to the use of unrealistically high viscosity.
Many studies have adopted a kinematic dynamo approach in which
the flow is prescribed for all time with no back-reaction from the magnetic field. 
The imposed mean base flow in these kinematic models can be chosen
analytically \citep{Kai99,Ava03,Gis08,Gis09,Gie10} or based on data from laboratory water
experiments \citep{Mar03,Rav05,Stef06,Lag08}. 
Some authors adopt a mean-field approach and  
parametrize the effects of small-scale turbulence through an $\alpha$-effect,
which corresponds to a mean electromotive force that is linear and homogeneous in 
the large-scale magnetic field
\citep{Ava03,Lag08,Gie10}.
A small number of studies use computationally expensive 3D self-consistent models where the velocity is 
produced by boundary or volume forcing, and the magnetic feedback on the flow is taken into account 
\citep{Bay07,Gis08b,Reu11}, but to our knowledge, only \citet{Rob10} have addressed the problem of magnetic boundary conditions
via self-consistent numerical simulations.

All previous numerical studies, using either the mean base flow only, the mean field approach,
or 3D self-consistent models, found that enhanced electrical conductivity or 
magnetic permeability of 
either the container walls or impellers 
lead to a reduction of the 
dynamo threshold measured by a critical magnetic Reynolds number
(where the magnetic Reynolds number corresponds to
the ratio of magnetic induction over magnetic diffusion).
\citet{Ava03} attribute the reduction of the dynamo threshold 
to a change in geometry of the electric current lines or the
magnetic field lines leading to a reduction of the total ohmic dissipation. 
\citet{Gie10} alternatively 
invoke the reduction of the ``effective'' magnetic diffusivity, that is the magnetic diffusivity averaged over the whole volume of
the system, although they acknowledge that this argument does not explain why different magnetic
field growth rates are obtained when 
varying individually either the magnetic permeability or the electrical conductivity of the disks.
\citet{Pet07} argue that the refraction of the magnetic field lines 
in the soft iron disks (due to the discontinuity of the tangential magnetic field)
may act as a shield 
for the fluid dynamo region between the two disks from the region behind the disks.
Indeed, \citet{Stef06} have shown that the motions of liquid sodium in the region behind the disks
is detrimental for the dynamo action in kinematic simulations.
\citet{Rob10} show that finite values of the wall conductance promote dynamo action even when
the wall permeability tends to zero. When the wall conductance tends to zero, on the other hand,
their model fails to produce a dynamo even for infinite wall permeability.
\citet{Kai99} find that a surrounding wall of the same conductivity as the fluid
favors dynamo action up to an optimal thickness. In this case, the ohmic dissipation in the fluid decreases
as the electric currents diffuse into the wall. 
However, they show that thicker walls are detrimental to dynamos that produce time-dependent magnetic field
because the skin effect leads to the presence of eddy currents in the wall,
and so the ohmic dissipation increases in this case.
In an experimental setup similar to VKS but using gallium as working fluid (which has a lower conductivity
than sodium and thus a lower magnetic Reynolds number) and applying 
transverse magnetic fields, \citet{Ver10} find that the induced axial magnetic field 
measured in the mid equatorial plane
is two to three times larger in magnitude when soft iron disks are used compared to copper or
stainless steel disks. They argue that this result (among other observations) is consistent 
with an induction mechanism in the rotating disks amplified by the distortion of the magnetic field lines
by the soft iron.

In general, in boundary-driven system, some essential component of the dynamo likely operates close to the
boundaries, so it is perhaps not surprising that changing the magnetic 
boundary conditions significantly affects the dynamo threshold. 
Nevertheless, the current physical interpretations of the experimental observations 
are partly based on assumptions about the flow properties in kinematic dynamo models,
and have not been demonstrated in self-consistent models.
Moreover, the cylindrical geometry of the VKS experiment 
makes it difficult to implement realistic boundary conditions numerically, which has led
some authors to adopt idealized boundary conditions (\emph{e.g.} infinite magnetic permeability
which implies vanishing tangential magnetic field at the boundary) 
\citep{Gis08,Lag08}.

Here we investigate the underlying problem 
of the role of magnetic boundary conditions in dynamo models
through self-consistent three-dimensional magnetohydrodynamical 
numerical simulations in spherical shell geometry.
Our study extends and expands on the work of \citet*{Rob10} (RGC10 hereafter)
who used
the boundary forcing exerted by the counter-rotation of the two hemispheric outer walls to drive 
a mean flow in a spherical cavity. 
In their study, RGC10 use a thin wall boundary condition which implies that the magnetic field
in the outer wall instantly responds to a change of magnetic field in the fluid. 
However the conditions under which the thin wall limit is appropriate for modeling
the experimental setup are unclear.
Here, we investigate in more details the role of the outer wall by modeling a wall of
finite thickness and finite values for the electrical conductivity and magnetic permeability.
The model is self-consistent in the sense that the flow produced by the motions of the rotating
boundaries can be adjusted by the Lorentz forces of the sustained magnetic field. 
Furthermore, no parameterization of the turbulent effects are included in the equations.
For fixed forcing and magnetic properties of the fluid, we have examined
the effects of varying the properties of the wall
(magnetic permeability $\mu_w$, electrical conductivity $\sigma_w$, and thickness $h$)
on the resultant dynamo. 
Spherical geometry has the advantage that magnetic boundary 
conditions can be easily implemented using a toroidal-poloidal decomposition for the 
magnetic field.  
The spherical geometry is convenient to study magnetohydrodynamical (MHD) problems numerically
but prevents us from studying the exact same flow obtained in the cylindrical VKS experiment.
Moreover, the impellers in the VKS experiment consist of flat disks upon which
eight curved blades are attached. The effects of the blades on the flow are not reproduced in our numerical
setup. Therefore the application of our results to the VKS experiment will remain tentative. 

\section{Model}

\begin{figure}
\centering
   \includegraphics[clip=true,width=0.6\textwidth]{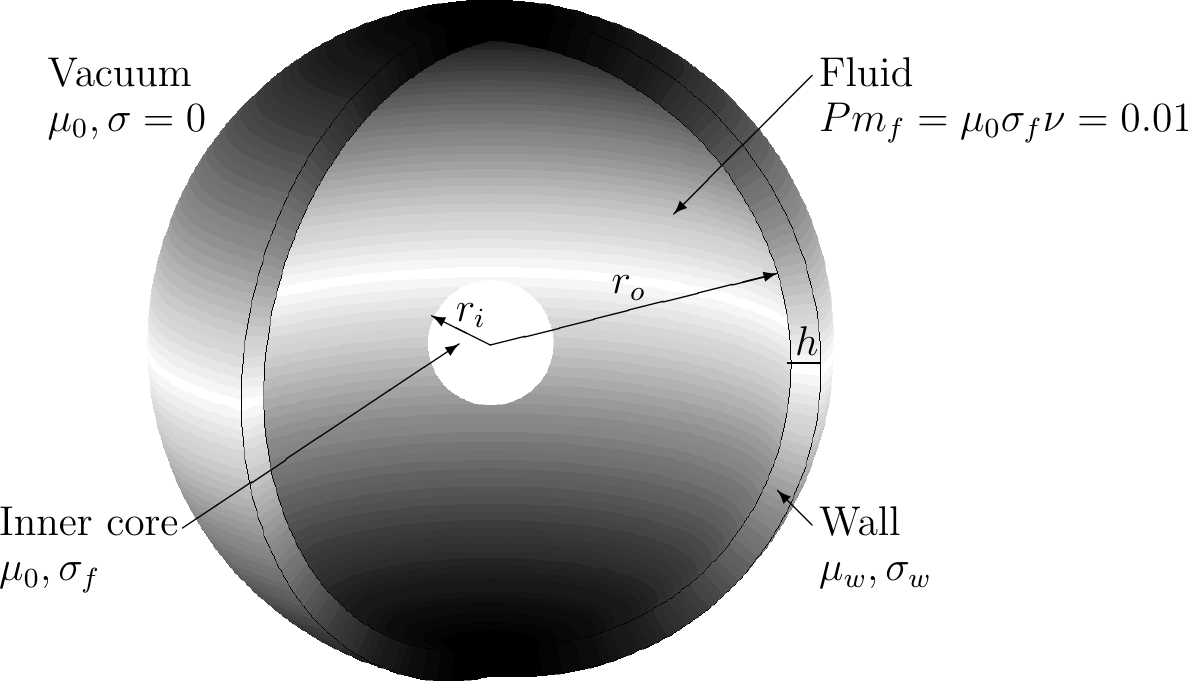}
   \caption{Schematic 3D view of the setup. Within the wall, the shading represents the absolute value of the angular velocity 
   $| \Omega_w |=U_w |\cos \theta| /r_o$, maximum at the poles and zero at the equator.}
\label{fig:schema_setup}
\end{figure}

The model setup is sketched in Figure~\ref{fig:schema_setup}. 
We use spherical coordinates $(r, \theta, \phi)$ 
with $r$ the radius, $\theta$ the colatitude and $\phi$ the azimuth.
An electrically conducting incompressible fluid fills the spherical shell between the inner radius
$r_i$ and the outer radius $r_o$. The fluid has viscosity $\nu$,
density $\rho$,
electrical conductivity $\sigma_f$, and its magnetic permeability is equal
to the vacuum magnetic permeability, $\mu_0$. 
All the properties of the fluid are constant in the volume and are not varied 
throughout the MHD study.
The solid outer wall is modelled by a spherical shell of finite thickness $h$
which rotates with angular velocity 
\begin{eqnarray}
	\Omega_w(\theta)=U_w \cos \theta /r_o ,
\label{eq:Omega_wall}
\end{eqnarray}
where $U_w$ is a constant forcing velocity.
Northern and Southern hemispheres therefore rotate at the same rate but in opposite directions.
Note that the wall is  ``solid'' in the sense that it is not fluid, but the angular velocity profile~(\ref{eq:Omega_wall}) varies with latitude
and so a shear is present in the wall.
We impose impenetrable and no-slip boundary conditions on the fluid at $r = r_o$, and so the wall exerts a viscous stress on the fluid.
The wall has electrical conductivity $\sigma_w$ and
magnetic permeability $\mu_w$.
At the outer boundary of the wall, $r=r_o+h$, we impose a vacuum boundary condition
corresponding to zero electric current for $r > r_o + h$.

Within the fluid, we solve the momentum equation for an incompressible fluid and the magnetic induction equation:
\begin{eqnarray}
	&& \pdt{\vect{u}} + \vect{u} \cdot \nabla \vect{u} = -\frac{1}{\rho} \nabla p + \nu \nabla^2 \vect{u} + \frac{1}{\rho} \vect{j} \times \vect{B},
	\\
	&& \nabla \cdot \vect{u} = 0,
	\\
	&&\pdt{\vect{B}} = \nabla \times \pleft \vect{u} \times \vect{B} \pright - \nabla \times \frac{1}{\sigma_f} \nabla \times \frac{\vect{B}}{\mu_0},
	\\
	&&\nabla \cdot \vect{B} = 0 ,
\end{eqnarray}
where $\vect{u}$ is the velocity, $p$ is the pressure, and $\vect{B}$ is the magnetic field.
Within the wall, we solve the magnetic induction equation 
with only the prescribed velocity for the wall \mbox{$u_{\phi}(\theta,r)=\Omega_w(\theta) r \sin \theta$}
where $\Omega_w$ is given by Equation~(\ref{eq:Omega_wall}).
More details about the implementation of the induction in the wall are given in Appendix~\ref{app:BC}.
For numerical convenience a solid inner core is present between \mbox{$r=0$} and \mbox{$r=r_i=0.2r_o$}.
The inner core is held at rest and has the same electrical conductivity and magnetic permeability as the fluid. 
We solve the magnetic induction equation within the inner core 
with zero velocity. The boundary conditions for the velocity at \mbox{$r=r_i$} are no-slip and impenetrable.

The equations are solved in non-dimensional form. 
The length scale is the outer radius $r_o$; the velocity is scaled by the forcing velocity $U_w$;
the time is scaled by $r_o/U_w$;
the magnetic field is scaled by \mbox{$\sqrt{\rho \mu_0} U_w$}.
The dimensionless parameters for the fluid are the magnetic Prandtl number:
\begin{eqnarray}
	\Pm_f=\mu_0 \sigma_f \nu,
\end{eqnarray}
and the Reynolds number, which is a measure of the forcing strength:
\begin{eqnarray}
	\Rey=\frac{U_{w} r_o}{\nu}.
\end{eqnarray}
The dimensionless parameters for the wall are the wall thickness \mbox{$\hat{h}=h/r_o$}, the relative conductivity
\mbox{$\sigma_r=\sigma_w/\sigma_f$}, and the relative magnetic permeability \mbox{$\mu_r=\mu_w/\mu_0$}.

At the fluid-wall interface, the boundary conditions 
for the normal and tangential components of the magnetic field 
and electric current density, \mbox{$\vect{j}=\nabla \times \mu^{-1}\vect{B}$}, are:
\begin{eqnarray}
	&& \vect{B}_w \cdot \vect{e}_r = \vect{B}_f \cdot \vect{e}_r,
	\label{eq:Bnormal}
	\\
	&& \vect{B}_w \times \vect{e}_r = \mu_r \vect{B}_f \times \vect{e}_r,
	\label{eq:Btan}
	\\
	&& \vect{j}_w \cdot \vect{e}_r = \vect{j}_f \cdot \vect{e}_r,
	\label{eq:jnormal}
	\\
	&& \vect{j}_w \times \vect{e}_r = \sigma_r \vect{j}_f \times \vect{e}_r,
	\label{eq:jtan}
\end{eqnarray}
where the subscripts $f$ and $w$ indicate the field on the side of the fluid and wall respectively.
Therefore a jump of magnetic permeability, \mbox{$\mu_r \ne 1$}, implies a discontinuity of the tangential magnetic field
whereas a jump of electrical conductivity, \mbox{$\sigma_r \ne 1$}, implies a discontinuity of the tangential electric currents, 
and hence a discontinuity in the radial derivatives of the tangential magnetic components.

For this study, we have modified the fully three-dimensional and non-linear PARODY code 
that was designed to solve magnetohydrodynamic problems in spherical geometry.
The modification includes the addition of an outer wall of finite thickness and finite magnetic properties.
The code was originally written by \citet{Dor98},
and subsequently parallelized and optimized by J.~Aubert and E.~Dormy.
The code was previously benchmarked against 5 independent numerical codes used in the geophysical and astrophysical 
dynamo community \citep{Chr01}.
The velocity and magnetic fields are decomposed into poloidal and toroidal scalars, 
which are then expanded in spherical harmonics $Y_l^m$ in the angular coordinates
with $l$ representing the latitudinal degree and $m$ the azimuthal order.
A second-order finite difference scheme is used on an irregular radial grid (finer near the boundaries, using
a geometrical progression for the radial increment). 
A Crank-Nicolson scheme is implemented for the time integration of the diffusion terms and 
an Adams-Bashforth procedure is used for the other terms.
The poloidal-toroidal decomposition and the spherical geometry allow a relatively simple 
implementation of the magnetic boundary conditions.
A detailed description of the implementation of the magnetic boundary conditions in the code is given in Appendix~\ref{app:BC}.

The typical resolution is $400$ radial points in the fluid, between $20$ and $50$ radial points in the wall depending on the
wall parameters, $15$ radial points in the inner core, \mbox{$l_{\textrm{max}}=160$} degrees and \mbox{$m_{\textrm{max}}=48$} orders of spherical harmonics. 
We have verified that the kinetic and magnetic energy spectra in $l$ and $m$ are well resolved 
(see Figures~\ref{fig:Spectrum_KE_Re48193}, \ref{fig:spectre_ME} and \ref{fig:spectre_mag_m}),
and that a finer radial resolution does not change the numerical solution significantly.
All the simulations presented in this paper have reached a statistically stationary state in which the
kinetic and magnetic energies are roughly constant in time.

The relations between spherical components and poloidal-toroidal components are given in Appendix~\ref{app:BC}.
In the following sections, poloidal and toroidal scalars of a vector $\vect{B}$ are denoted $B_P$ and $B_T$ and
the radial, latitudinal and azimuthal components $B_r$, $B_{\theta}$ and $B_{\phi}$ respectively.
For an axisymmetric field (corresponding to $m=0$ in spectral space) 
the toroidal component is related to the azimuthal component, directed East-West,
and the poloidal component is related to the radial and latitudinal components, enclosed in a meridional plane.
Azimuthal averages are denoted by an overbar.

\section{Results}
First, in Section~\ref{sec:hydro}, we present hydrodynamic simulations without magnetic field for different
values of the boundary forcing in order to study the underlying flow in the system. 
The rest of the paper then focuses on the results from full MHD simulations run at a fixed forcing.
Sections~\ref{sec:onset} and~\ref{sec:char_MF} describe the dynamo onset and some general 
features of the self-sustained magnetic field. 
In Section~\ref{sec:dynamo_mec}, we investigate the details of the dynamo process
and discuss the role of the magnetic properties of the wall.

\subsection{\label{sec:hydro} Hydrodynamics}

\begin{figure*}
\centering
   \subfigure[$\Rey=300$]{\label{fig:mean_Re300}
   \includegraphics[clip=true,width=5.5cm]{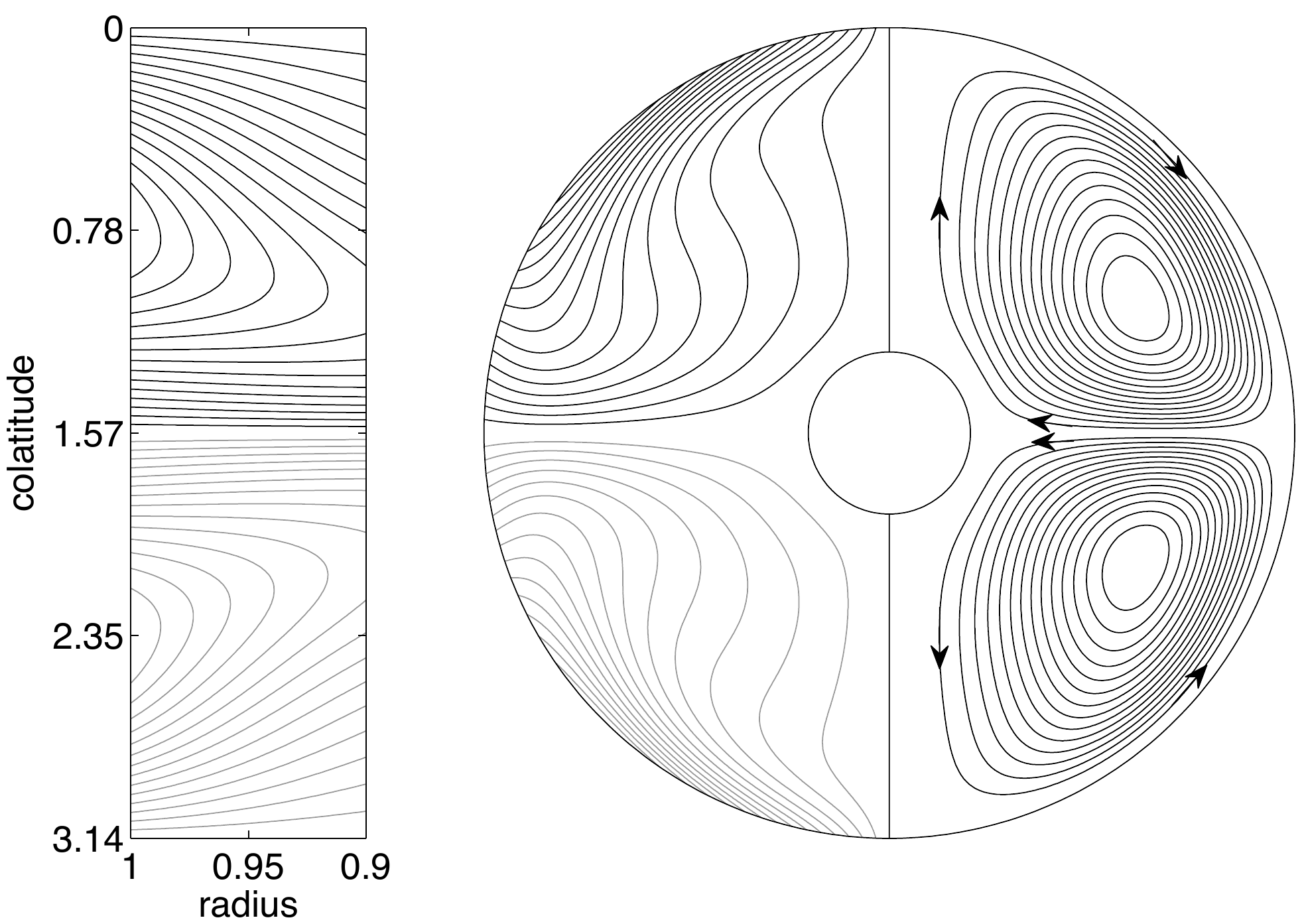}}
   \subfigure[$\Rey=48193$]{\label{fig:mean_Re48193}
   \includegraphics[clip=true,width=5.5cm]{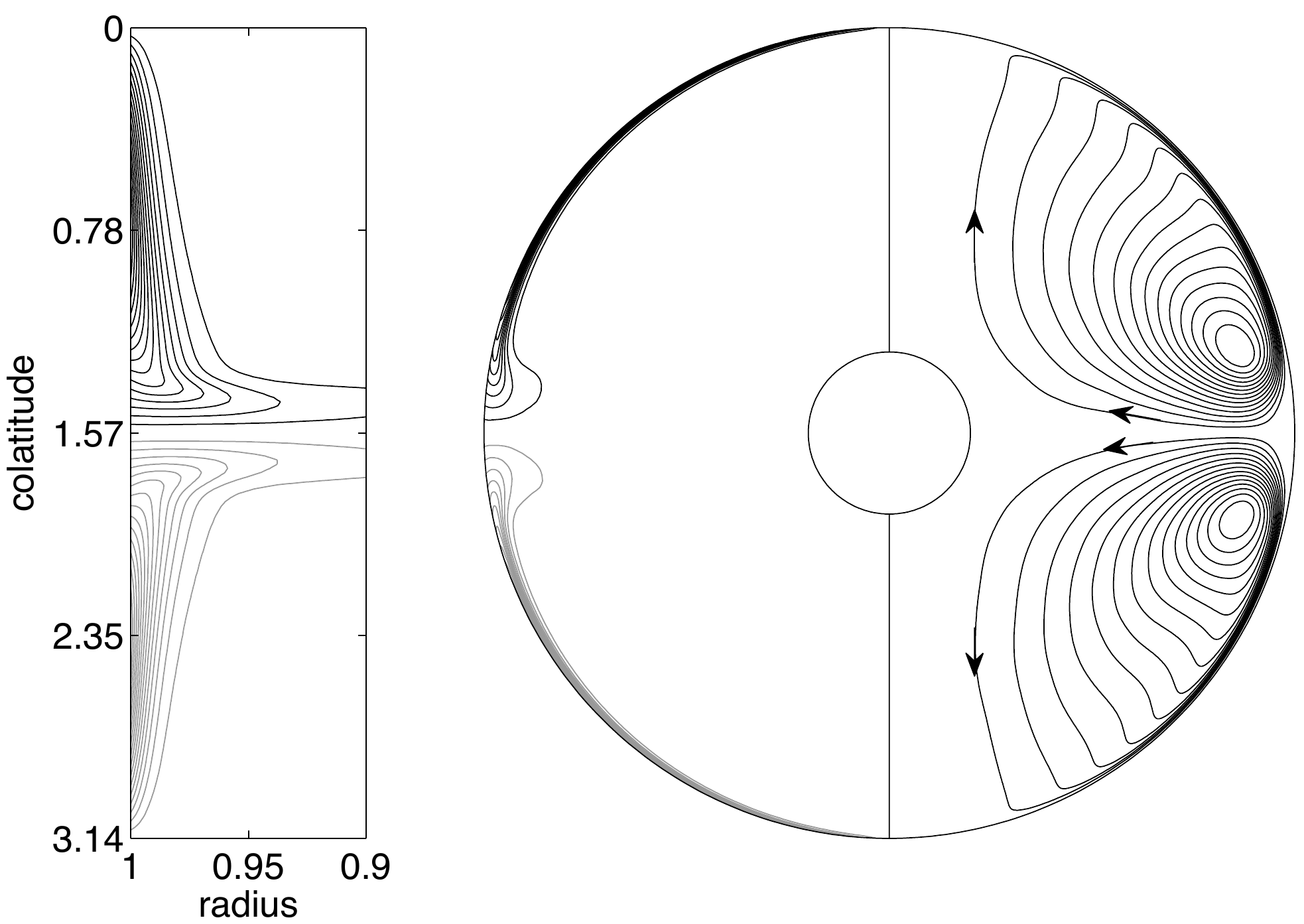}}
\caption{Axisymmetric flow in a meridional plane for different Reynolds numbers in the statistically steady state. 
The left panel (close-up of the outer boundary layer) and left half of the
 meridional plane show the zonal velocity $\overline{u_{\phi}}$ (black: positive; gray: negative).
The right half of the meridional plane shows the axisymmetric poloidal streamlines 
(corresponding to \mbox{$\overline{\vect{u}}_{pol}=(\overline{u_r},\overline{u_{\theta}})$}) with constant contour interval. 
Extremum values for the axisymmetric velocity are (in units of $U_w$): 
for \mbox{$\Rey=300$}: \mbox{$\overline{u_r} \in [-0.095,0.090]$},
\mbox{$\overline{u_{\theta}} \in [-0.09,0.09]$}, and \mbox{$\overline{u_{\phi}} \in [-0.50,0.50]$};
for \mbox{$\Rey=48193$}, \mbox{$\overline{u_r} \in [-0.023,0.011]$},
\mbox{$\overline{u_{\theta}} \in [-0.10,0.10]$}, and \mbox{$\overline{u_{\phi}} \in [-0.50,0.50]$}.}
\label{fig:flow_mean}
\end{figure*}
\setcounter{subfigure}{0}

The axisymmetric viscous forcing exerted by the outer boundary on the fluid drives a
zonal (\emph{i.e.} axisymmetric and azimuthal) velocity, $\overline{u_{\phi}}$.
For small Reynolds numbers (\emph{e.g.} \mbox{$\Rey=300$} in Figure~\ref{fig:mean_Re300}),
the zonal velocity extends into the bulk of the fluid whereas for large $\Rey$
(\emph{e.g} \mbox{$\Rey=48193$} in Figure~\ref{fig:mean_Re48193}), the zonal flow is
confined to a narrow layer close to the outer boundary. 
Within this layer at large $\Rey$, radial gradients of $\overline{u_{\phi}}$ are large 
at all latitudes except in the equatorial region, where latitudinal gradients are largest.
The differential rotation in the viscous boundary layer pumps a meridional (\emph{i.e.} axisymmetric and poloidal)
circulation consisting of two counter-rotating cells with
an inward radial velocity in the equatorial plane (Figure~\ref{fig:flow_mean}).
The viscous boundary layer is analogous to an Ekman boundary layer commonly found in rotating flows,
and the pumping of the meridional circulation is analogous to Ekman pumping \citep{Gre68}.
The thickness of the Ekman layer scales as \mbox{$\delta \propto \Rey^{-1/2}$}
in our dimensionless units.
Hence a thinner viscous boundary layer is observed for large Reynolds number.
The viscous timescale is \mbox{$\tau_{\nu}=\Rey$}, and  
the spin-up timescale is \mbox{$\tau_{\Omega}=\Rey^{1/2}$} in dimensionless units
\citep{Gre68}.
The spin-up timescale corresponds to the typical timescale to establish the meridional circulation, which transports angular momentum
between the viscous boundary layer and the bulk of the fluid. 
For large Reynolds numbers, or equivalently low
viscosity, the meridional circulation establishes a steady state much more rapidly than by viscous diffusion alone.
For example at \mbox{$\Rey=48193$}, the hydrodynamic simulation has been run for about 4 $\tau_{\Omega}$, corresponding to a few percent of 
a viscous timescale. After a short initial transition phase (lasting about 50 time units), the kinetic energy reaches a constant average value
and no further spreading of the viscous boundary layer is observed.
Since the upper and lower hemisphere rotates at opposite rotation rate,
we expect the bulk of the fluid to have zero angular velocity as observed for \mbox{$\Rey=48193$}. 

\begin{figure*}
\centering
   \subfigure[$\Rey=300$ (snapshot)]{\label{fig:Una3d_Re300}
   \includegraphics[clip=true,height=4cm]{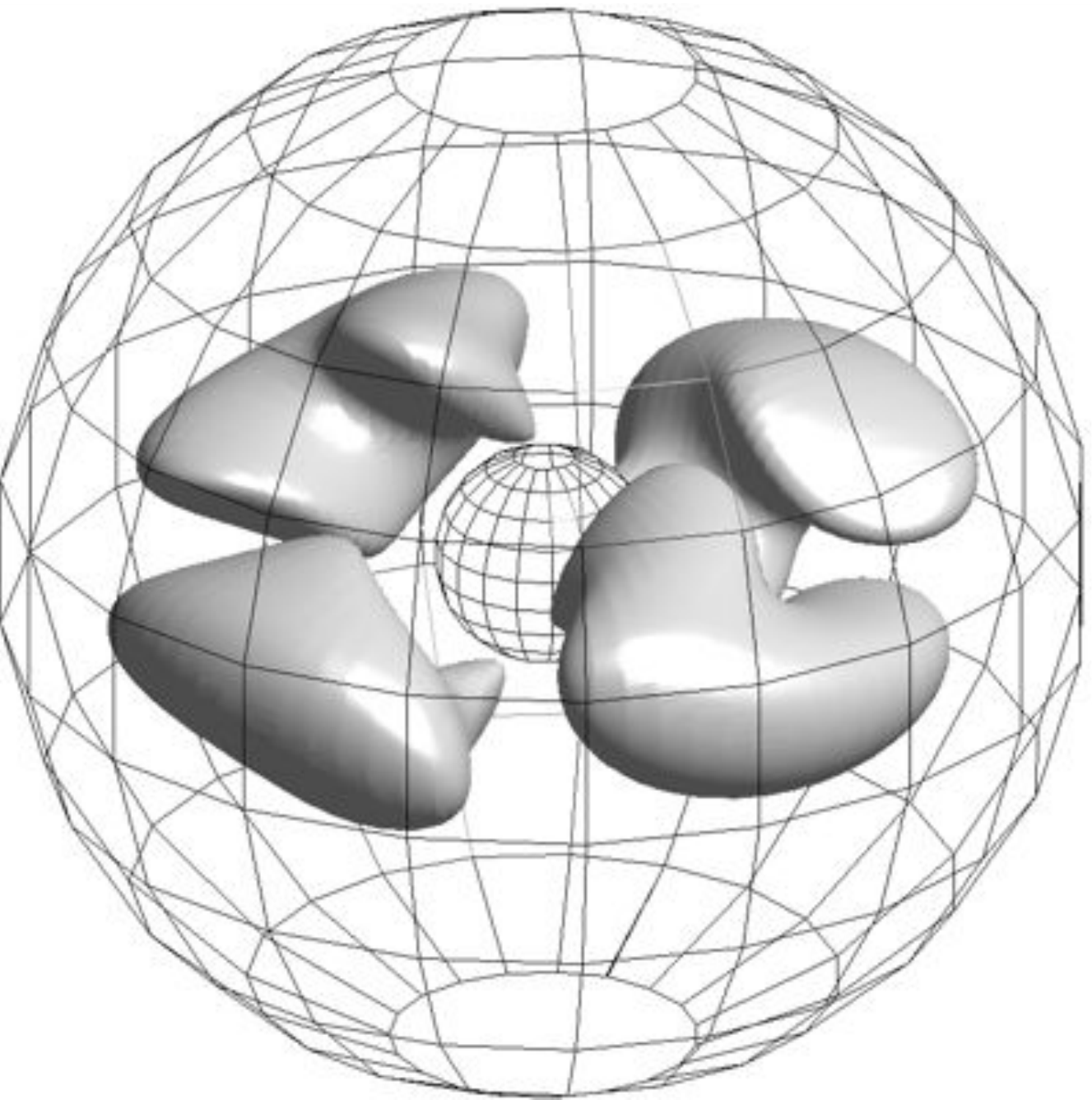}}
   \subfigure[$\Rey=48193$ (snapshot)]{\label{fig:Una3d_Re48193}
   \includegraphics[clip=true,height=4cm]{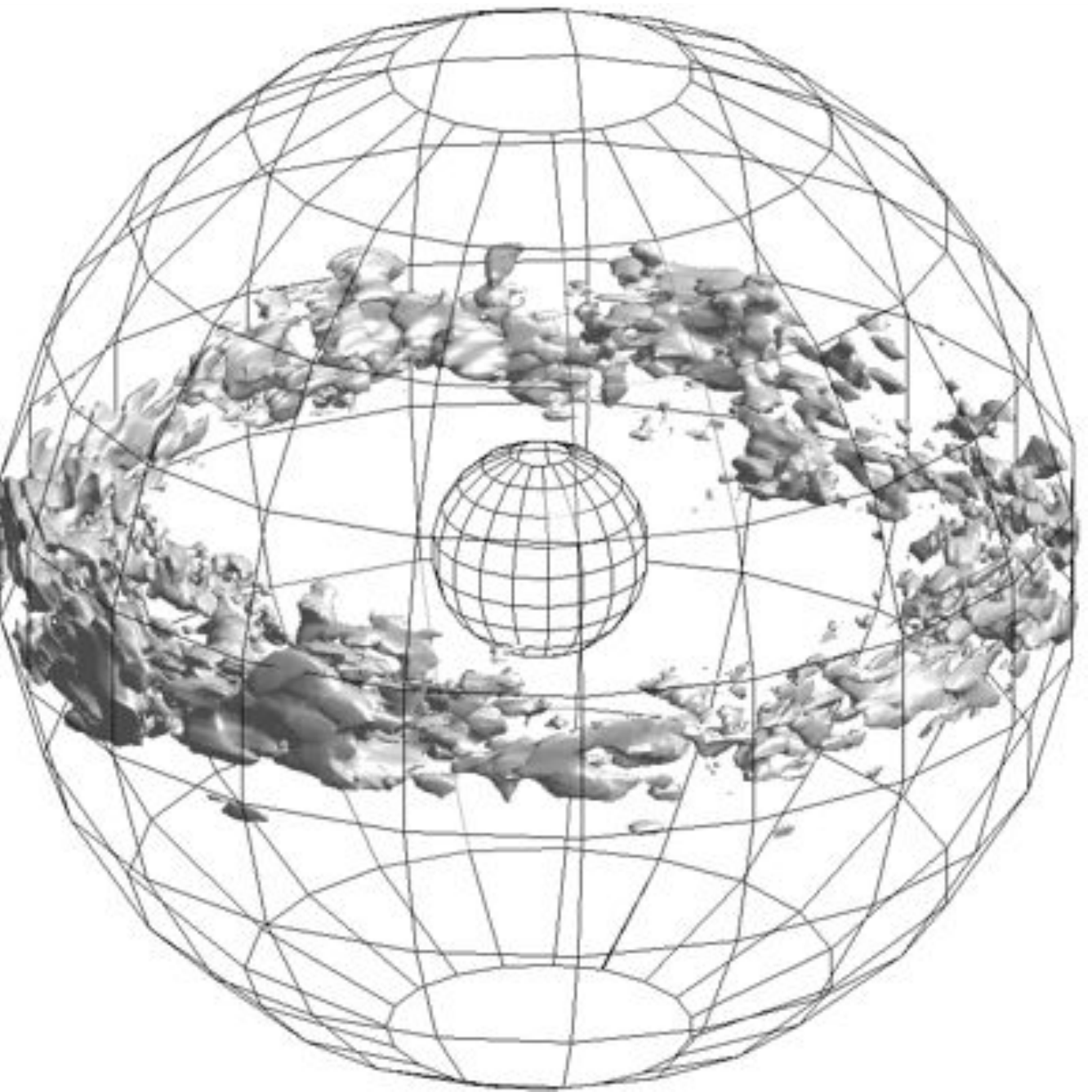}}
    \subfigure[$\Rey=48193$ (time-average)]{\label{fig:Una_Re48193}
    \raisebox{0.2cm}{\includegraphics[clip=true,height=4cm]{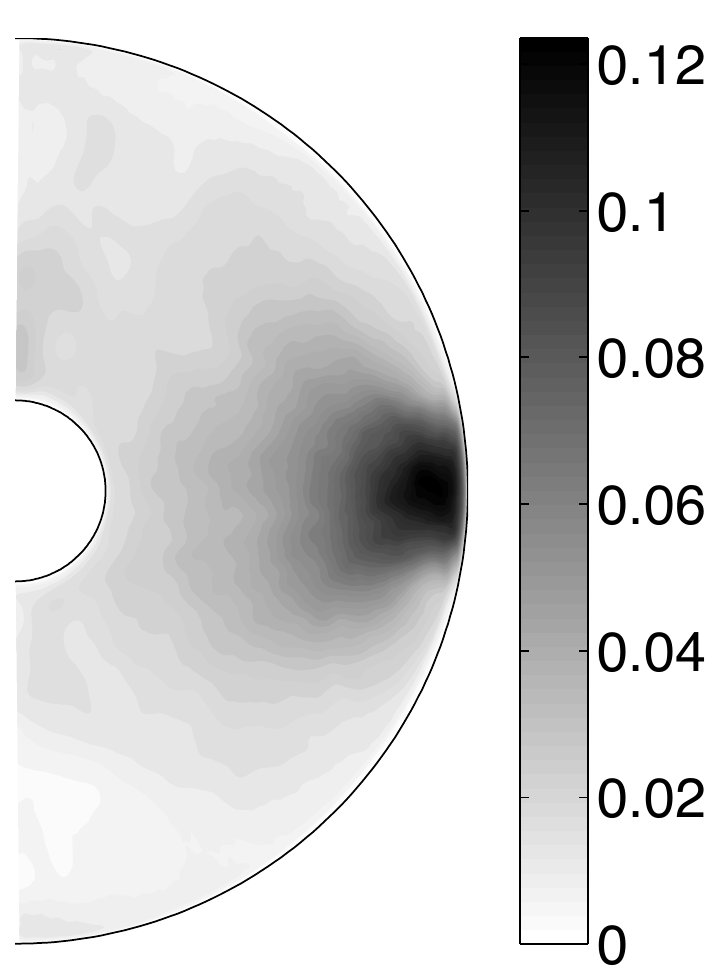}}}
    \subfigure[Kinetic energy spectrum $\Rey=48193$]{\label{fig:Spectrum_KE_Re48193}
    \includegraphics[clip=true,height=4cm]{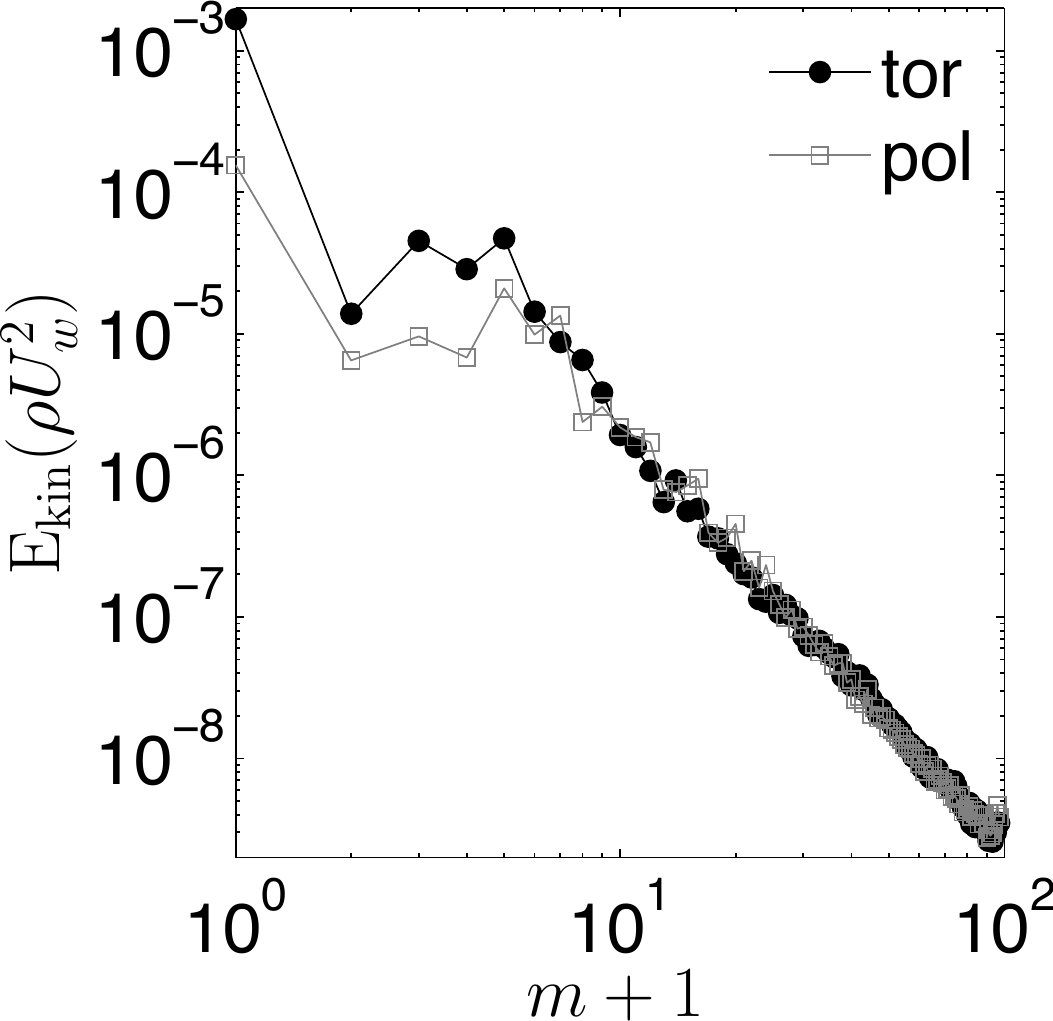}}
\caption{(a) and (b) Isodensity surface of kinetic energy of the non-axisymmetric components (20\% ($\Rey=300$) and 15\% ($\Rey=48193$) of the maximum). 
(c) Azimuthal and time average of the non-axisymmetric rms velocity ($\sqrt{(|\vect{u}|-|\overline{\vect{u}}|)^2}$) for $\Rey=48193$ in units of $U_w$.
(d) Kinetic energy spectrum for azimuthal modes in the toroidal component (black circles) and in the poloidal component (gray squares)
 for $\Rey=48193$. The kinetic energy has been 
 averaged in time and over the whole volume.}
\end{figure*}
\setcounter{subfigure}{0}
 
The latitudinal profile of $\overline{u_{\phi}}$ imposed at the boundary displays an inflection
point at the equator. Consequently, for small enough viscosity, the steady axisymmetric base flow is prone to
shear instabilities according to the Rayleigh instability criterion for shear flows \citep[\emph{e.g.}][]{Dra66}.
Based on numerical simulations, we have determined that a non-axisymmetric 
component of the flow first appears at a critical Reynolds number \mbox{$200<\Rey_c<250$}.
The most unstable mode at \mbox{$\Rey=250$} has the azimuthal symmetry \mbox{$m=2$} and consists 
of vortices located around the equatorial plane (Figure~\ref{fig:Una3d_Re300}).
In a similar von K\'arm\'an flow in cylindrical geometry,  
\citet{Nor03} argued that the first non-axisymmetric instability of the 
equatorial shear layer of the basic flow is similar to a Kelvin-Helmoltz instability. 

For \mbox{$\Rey=48193$}, which is about 200 times larger than the critical Reynolds number, the non-axisymmetric flow
is still mostly located in the outer equatorial region, which we call the equatorial belt (Figures~\ref{fig:Una3d_Re48193} and~\ref{fig:Una_Re48193}).
On average, the energy of the non-axisymmetric flow is symmetric with respect to the equatorial plane.
The mean non-axisymmetric velocity tends to zero on a long time average. 
The mean flow is therefore purely axisymmetric.
For this Reynolds number, kinetic energy is found in all azimuthal modes, but the 
\mbox{$m=2, 4, 6$} modes have significantly larger amplitude than the other non-axisymmetric modes
(Figure~\ref{fig:Spectrum_KE_Re48193}).
For \mbox{$m \geqslant 8$} the spectrum approximately follows a $m^{-3}$ power law.
Note that \mbox{$k_{\phi}=m/(r \sin \theta)$} corresponds to the azimuthal wavenumber and 
\mbox{$k_H=\sqrt{l(l+1)}/r \approx (l+1/2)/r$} is the horizontal wavenumber on a spherical surface \citep{Bac96}.
In our numerical simulation at \mbox{$\Rey=48193$}, the kinetic energy spectrum in $l$ also follows a $l^{-3}$ power law at small scales.  
This result is somewhat unexpected for 3D turbulence at small scales in these types of flows, 
for which the latest theoretical predictions and experimental results (albeit performed in cylindrical 
geometry) obtain spatial kinetic spectra exhibiting a $k^{-2}$ slope
when the cascade is non-local at small scales \citep{Her12}. 
A $k^{-3}$ spectrum is more generally associated with quasi-2D turbulence, 
which could speculatively be argued here but is not obviously the case.

We define a local Reynolds number, \mbox{$\Rey_l=(u^{\ast}\pi/m) \Rey$} where $u^{\ast}(m)$ is the 
dimensionless rms velocity for a mode $m$, as a measure of the local
ratio of the non-linear inertial terms to the viscous terms. 
The viscous scale, defined as the scale for which \mbox{$\Rey_l\approx1$}, is \mbox{$m=43$}.
For the MHD simulations presented in the next sections, the magnetic diffusive scale, defined for a local magnetic
Reynolds number of the conducting fluid of order unity, \mbox{$\Rm_l=\Pm_f \Rey_l=1$}, is
\mbox{$m=5$} (\mbox{$\Pm_f=0.01$}).

 \begin{figure*}
\centering
    \subfigure[]{\label{fig:EK_Re}
    \includegraphics[clip=true,width=5.8cm]{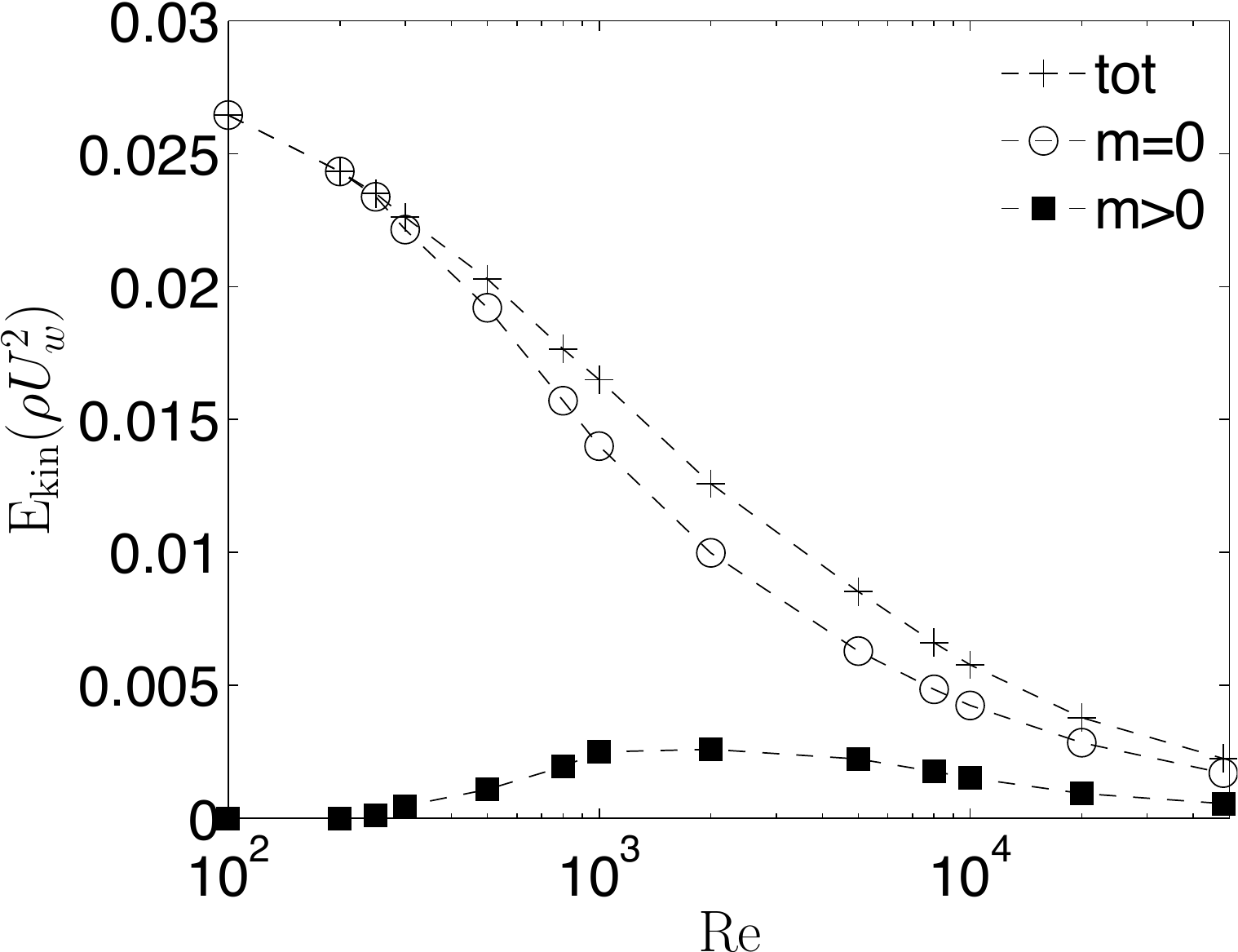}}
    \subfigure[]{\label{fig:ratio_EK}
    \raisebox{0.1cm}{\includegraphics[clip=true,width=5.8cm]{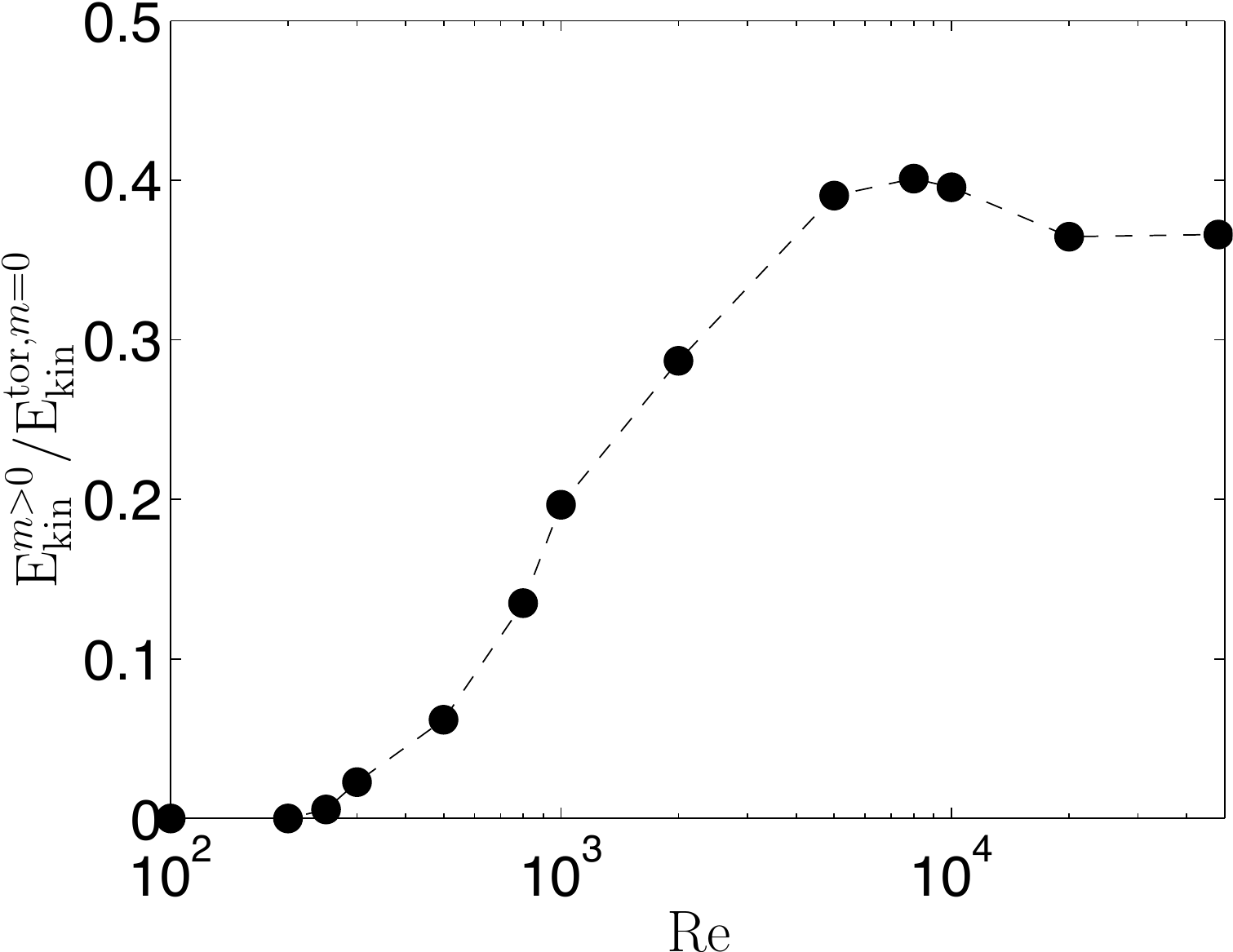}}}    
   \caption{(a) Kinetic energy contained in the total flow ($+$),
    in the axisymmetric flow ($\bigcirc$) and in the non-axisymmetric flow ($\blacksquare$) for different Reynolds numbers.
   (b) Ratio of non-axisymmetric to zonal kinetic energy for different Reynolds numbers.
   The kinetic energy is averaged in time and over the whole volume of the fluid.}
   \label{fig:Ekin_forcing}
\end{figure*}
\setcounter{subfigure}{0}

The mean kinetic energy as a function of the forcing Reynolds number is shown in Figure~\ref{fig:EK_Re}. 
The \emph{dimensionless} kinetic energy decreases as the Reynolds number increases, 
implying that either the laminar or the turbulent viscous dissipation increases. 
Note that the \emph{dimensional} kinetic energy increases with the forcing but the 
flow amplitude does not scale linearly with the forcing.
In Figure~\ref{fig:ratio_EK}, we plot the ratio of non-axisymmetric to 
zonal kinetic energies (volume-averaged) in function of $\Rey$. 
As expected, the ratio increases with the forcing for \mbox{$\Rey>\Rey_c$} but then saturates for \mbox{$\Rey>5000$}
at a value of about $37\%$.
We interpret this result as a consequence of the location of the non-axisymmetric motions in the equatorial belt,
where latitudinal gradients of $\overline{u_{\phi}}$ are large. 
The zonal kinetic energy remains mostly confined 
to the narrow laminar viscous boundary layer at higher latitudes (see Figure~\ref{fig:mean_Re48193}).
Therefore the non-axisymmetric flow only drains part of the zonal kinetic energy, even at large $\Rey$. 

\subsection{\label{sec:onset}Dynamo onset in the parameter space}
The MHD simulations
presented in this paper have been run with fixed Reynolds number $\Rey$ and fixed fluid properties
(magnetic permeability, electrical conductivity and viscosity).
We vary only the properties of the outer wall. 
To compare our results with RGC10, we use the same 
magnetic Prandtl number for the fluid, \mbox{$\Pm_f=\mu_0 \sigma_f \nu = 0.01$}.
Our forcing at \mbox{$\Rey=48193$}, about 200 times critical, corresponds to
\mbox{$U_w=80$m/s} with \mbox{$r_o=0.5$m} and \mbox{$\nu=8.3\times10^{-4}$m$^2$/s} in RGC10.
The magnetic Reynolds number of the fluid based on the forcing velocity 
at large scale is then \mbox{$\Rm_f=\Rey \Pm_f =482$}.
At these parameters, RGC10 obtained a dynamo 
if either the electrical conductivity and/or the magnetic permeability of the wall were made sufficiently large.
To investigate the role of the wall on dynamo action, we vary the wall's thickness $h$,
relative magnetic permeability $\mu_r$, and 
electrical conductivity $\sigma_r$.
The magnetic field is initialized with both a weak tilted dipole component
and a weak axial quadrupole component, with a ratio of magnetic energy
to kinetic energy of about $10^{-5}$.

\begin{figure*}
\centering
   \subfigure[$\hat{h}=0.1$]{\label{fig:dynspace_h.1}
   \includegraphics[clip=true,width=0.6\textwidth]{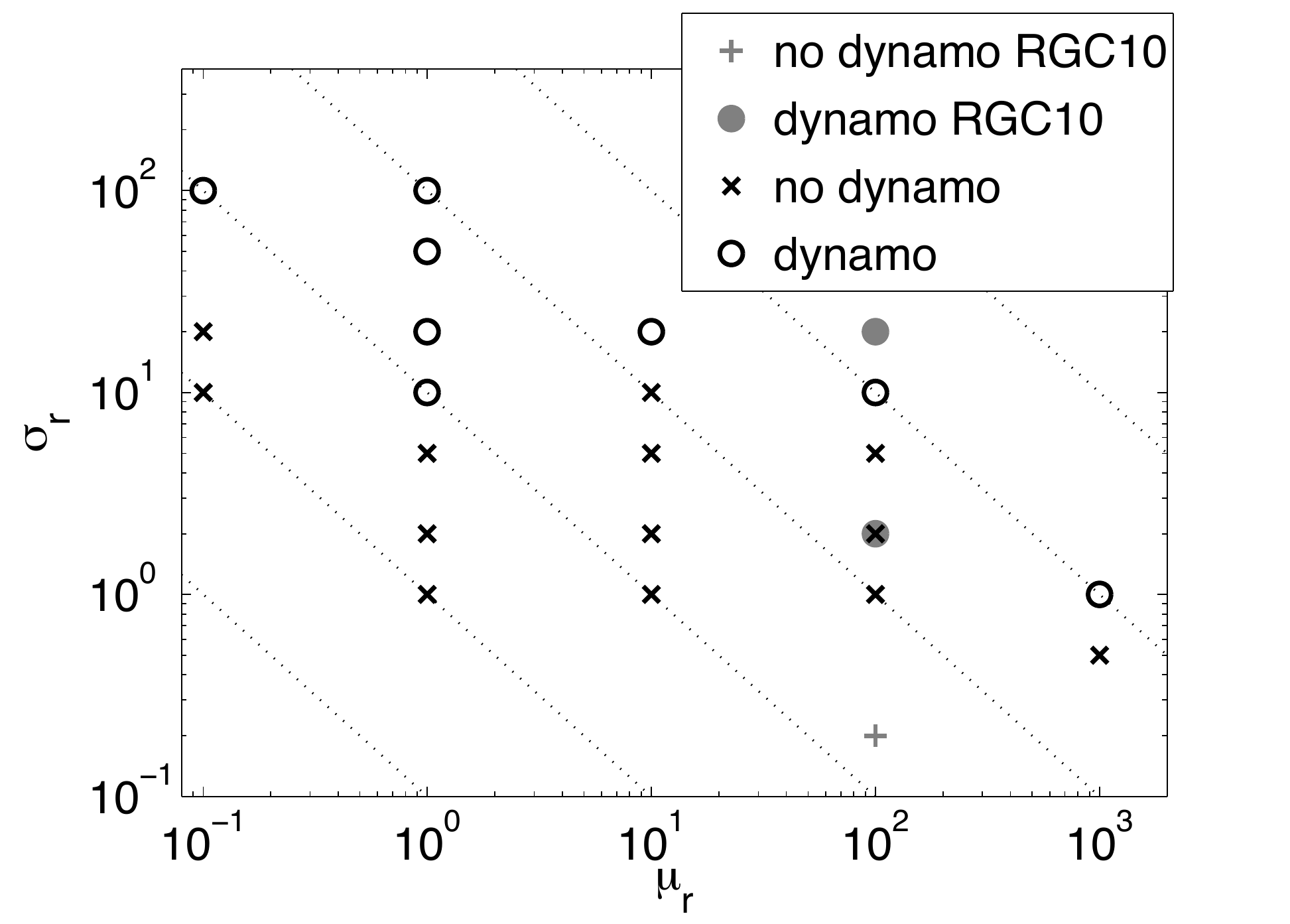}}
   \subfigure[$\hat{h}=0.01$]{\label{fig:dynspace_h.01}
   \includegraphics[clip=true,width=0.6\textwidth]{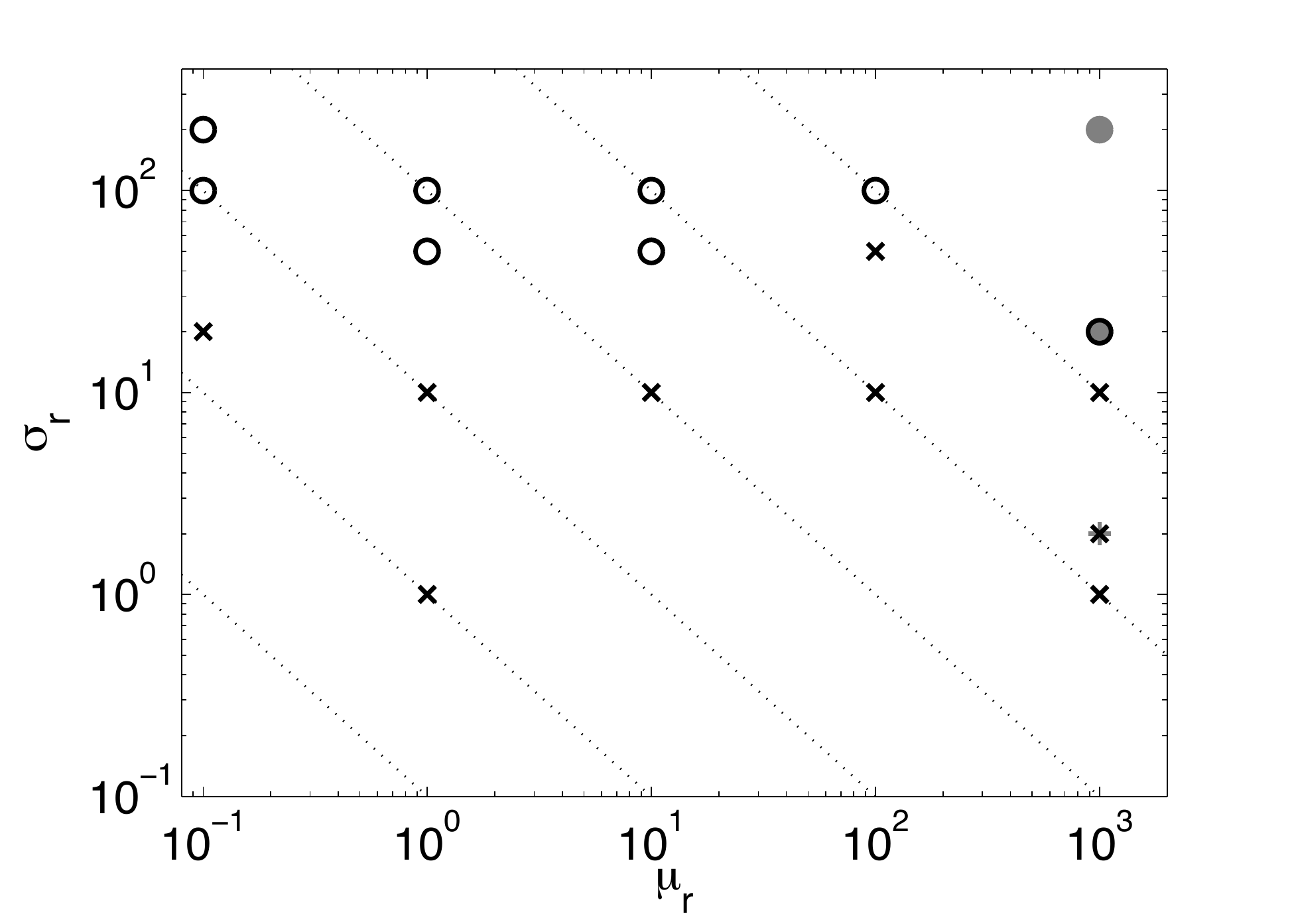}}
\caption{Results of MHD simulations in the parameter space ($\sigma_r$, $\mu_r$)
for two different wall thickness $\hat{h}$.
The black symbols are simulations presented in this study.
The gray symbols are simulations from RGC10. 
The dotted lines corresponds to constant magnetic diffusivities in the wall 
\mbox{$\eta_w=1/(\mu_w \sigma_w)$}.}
\label{fig:dynamo_space}
\end{figure*}
\setcounter{subfigure}{0}

Figure~\ref{fig:dynamo_space} presents the results in the parameter space \mbox{$(\sigma_r,\mu_r)$}
where $\sigma_r$ and $\mu_r$ are varied independently, and 
for two different wall thickness \mbox{$\hat{h}=0.1$} and \mbox{$\hat{h}=0.01$}.
We consider that a dynamo is in existence when the 
magnetic energy in the fluid stabilizes to a stationary value for several
global magnetic diffusion times in the fluid, \mbox{$\tau_f=\Rm_f$} in dimensionless units. 
When the global magnetic diffusion time of the wall \mbox{$\tau_w=\hat{h}^2 \sigma_r \mu_r \Rm_f$}
is larger than $\tau_f$, the simulation must be run for at least one time $\tau_w$.
For instance, the case \mbox{$\hat{h}=0.1$}, \mbox{$\sigma_r=1$} and \mbox{$\mu_r=1000$} has been run for
a time \mbox{$\tau_w=10\tau_f$}. 

For an homogeneous system (\mbox{$\mu_r=1$} and \mbox{$\sigma_r=1$}),
the flow is unable to generate a dynamo for \mbox{$\hat{h}=0.1$} and \mbox{$\hat{h}=0.01$}.

For \mbox{$\hat{h}=0.1$} (Figure~\ref{fig:dynspace_h.1}), an increase of the 
wall conductivity by a factor 10 while keeping \mbox{$\mu_r=1$} leads to dynamo action. 
An increase of the wall permeability by a factor 1000 is necessary to obtain dynamo
action while keeping \mbox{$\sigma_r=1$}.  
In this respect a large wall conductivity is more favorable for dynamo action than a  
large wall permeability. 
The boundary between dynamo and non-dynamo in the \mbox{$(\sigma_r,\mu_r)$} space 
does not follow a line of constant 
magnetic diffusivity of the wall, \mbox{$\eta_w=1/(\mu_w \sigma_w)$} (dotted lines in Figure~\ref{fig:dynamo_space});
therefore the effect of the wall on the dynamo mechanism cannot be understood purely in terms of magnetic diffusivity. 
Moreover, the loss of dynamo action from \mbox{$(\sigma_r, \mu_r) = (10, 1)$} to
\mbox{$(\sigma_r, \mu_r) = (10, 10)$} suggests that  
the effects of moderate wall conductivity and permeability act in competition and lead to
the dynamo suppression. 
The effects of $\sigma_r$ and $\mu_r$ on the dynamo process must therefore be considered separately.

For \mbox{$\hat{h}=0.01$} (Figure~\ref{fig:dynspace_h.01}), larger values of $\sigma_r$ and $\mu_r$ are necessary to obtain dynamo action
in comparison to \mbox{$\hat{h}=0.1$}. For \mbox{$\mu_r=1$}, an increase of the wall conductivity to \mbox{$\sigma_r=50$} is necessary 
to obtain a dynamo. 
For the large permeability case, \mbox{$\mu_r=1000$}, high values of the wall conductivity, at least \mbox{$\sigma_r=20$} are 
still required for dynamo action.
An increase of the wall thickness therefore appears to promote dynamo action (at least from \mbox{$\hat{h}=0.01$} to $0.1$).
Again, we observe competing effects between moderate values of both wall conductivity and permeability:
the case \mbox{$(\sigma_r,\mu_r)=(50,100)$} fails to produce a dynamo unlike the case \mbox{$(\sigma_r,\mu_r)=(50,10)$}.

In the thin-wall limit of RGC10, the controlling parameters are the
integrated conductivity over the wall thickness, \mbox{$\hat{h}\sigma_r$} 
and the integrated permeability \mbox{$\hat{h}\mu_r$}. Their results are shown in Figure~\ref{fig:dynamo_space}
with gray symbols.
For \mbox{$\hat{h}=0.1$}, the case \mbox{$(\sigma_r,\mu_r)=(2,100)$} fails to produce a dynamo in our study whereas
RGC10 obtained a dynamo.
However for \mbox{$\hat{h}=0.01$}, our results are in agreement with the results of RGC10,
implying that the thin-wall limit may be considered reasonably valid up to a wall thickness equal to 1\% of the outer radius but 
is not valid all the way up to 10\% relative thickness. 

We have re-run the dynamo case \mbox{$(\hat{h},\sigma_r,\mu_r)=(0.1,10,1)$}
as a kinematic dynamo.
We take the time-average of the velocity over $700$ 
rotation periods and then solve the induction equation with this prescribed velocity.
The flow is predominantly axisymmetric, but also contains weak non-axisymmetric components,
which are not coherent structures but rather the result of inadequate averaging. 
We find that this flow is unable to generate a dynamo,
proving that a key ingredient of the dynamo comes from the fluctuating part of the flow.

Finally, a brief search shows no evidence for subcritical dynamos. For instance, when the magnetic field from the case 
\mbox{$(\hat{h},\sigma_r,\mu_r)=(0.1,10,1)$}
is used as initial condition for the case \mbox{$(\hat{h},\sigma_r,\mu_r)=(0.1,1,1)$} the dynamo dies.

\subsection{\label{sec:char_MF}General characteristics of the magnetic field}

\begin{figure*}
\centering
   \subfigure[$(\sigma_r,\mu_r)=(1,1)$]{\label{fig:ME_sig1mu1}
   \includegraphics[clip=true,width=0.31\textwidth]{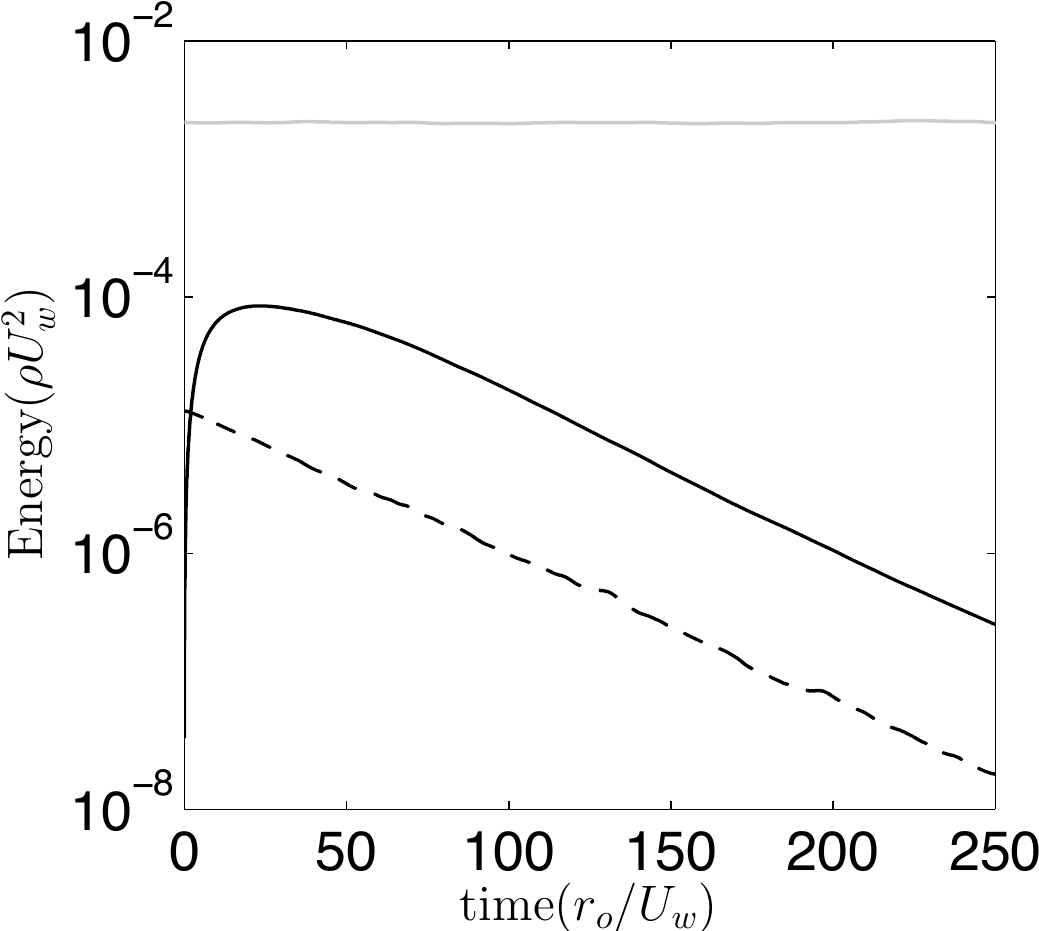}}
   \subfigure[$(\sigma_r,\mu_r)=(10,1)$]{\label{fig:ME_sig10mu1}
   \includegraphics[clip=true,width=0.31\textwidth]{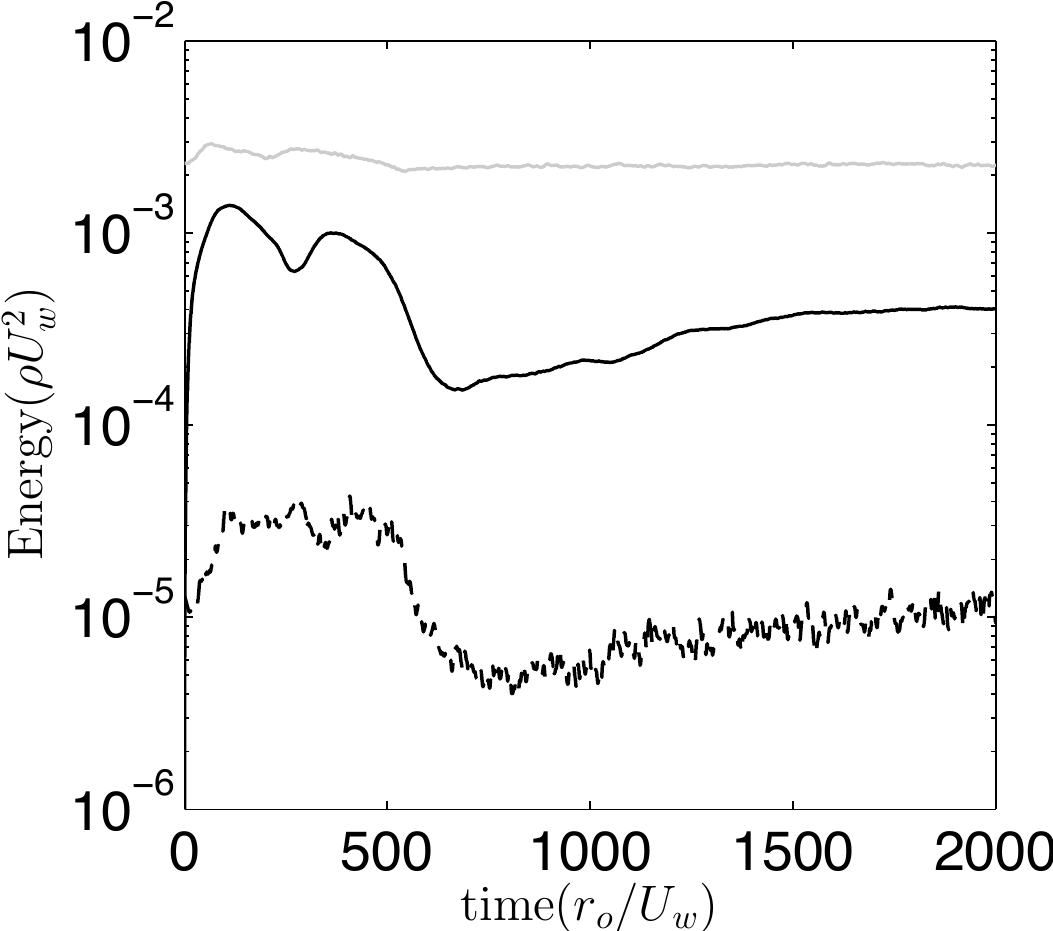}}
   \subfigure[$(\sigma_r,\mu_r)=(1,1000)$]{\label{fig:ME_sig1mu1000}
   \includegraphics[clip=true,width=0.31\textwidth]{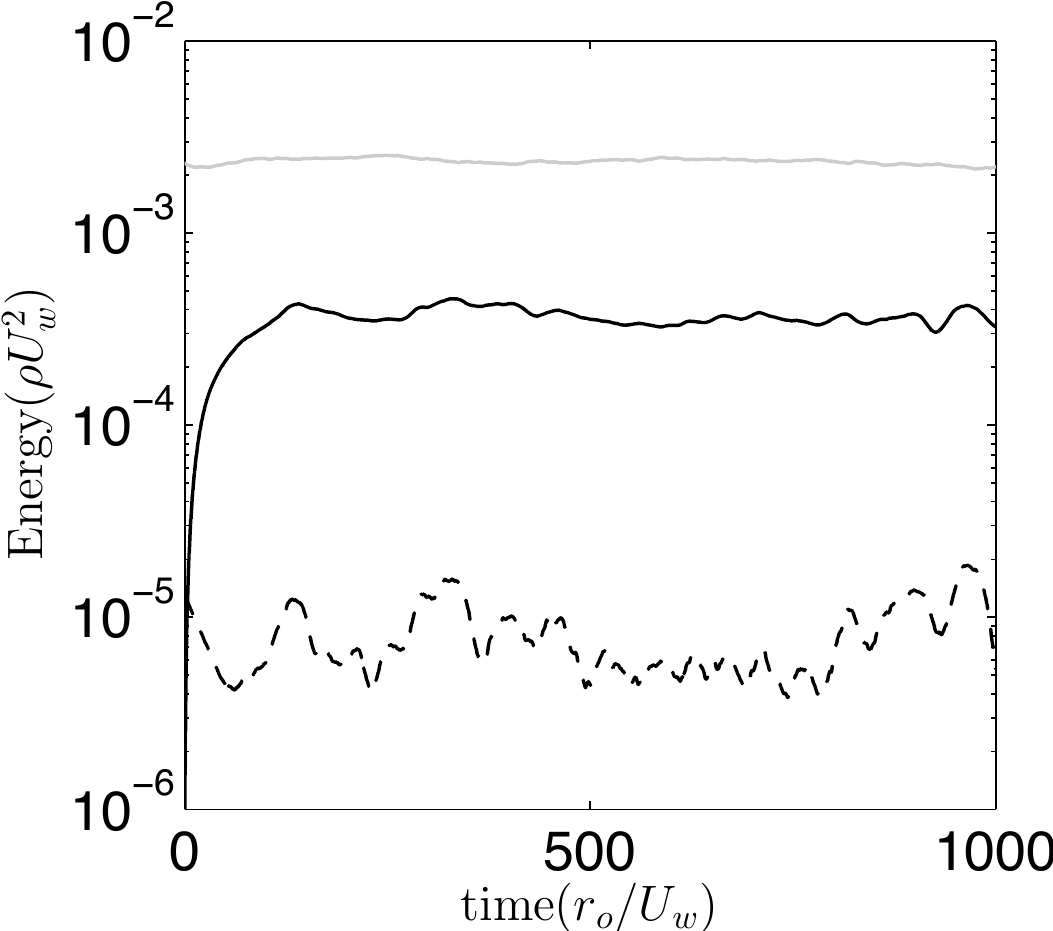}}
\caption{Time series of the kinetic (solid gray) and magnetic (toroidal: solid black and poloidal: dashed black) energies
for three simulations with different wall magnetic properties and $\hat{h}=0.1$. A global magnetic diffusion time
in the fluid $\tau_f$ corresponds to 482 time units. 
Kinetic and magnetic energy are averaged over the whole volume of the fluid.}
\label{fig:ME}
\end{figure*}

Figure~\ref{fig:ME} shows time series of the kinetic and magnetic energies of three cases -- a non-dynamo, 
\mbox{$(\hat{h},\sigma_r,\mu_r)=(0.1,1,1)$}, and two dynamos, $(0.1, 10, 1)$ and 
$(0.1, 1, 1000)$.
Close to the dynamo onset, the sustained magnetic energy is about 20\% 
of the averaged kinetic energy.
The most supercritical dynamo that we calculated (\mbox{$(\hat{h},\sigma_r,\mu_r)=(0.1,100,1)$})
produces an averaged magnetic energy about 30\% of the kinetic energy.
A comparison of energy spectra shows that the magnetic energy is smaller than the kinetic energy at all scales.
The kinetic energy value is not significantly modified in the saturated phase of the dynamo.
In the case \mbox{$(\hat{h},\sigma_r,\mu_r)=(0.1,10,1)$}, the kinetic energy increases initially as the magnetic field grows
to larger values than its saturated value (\mbox{$t<500$} in Figure~\ref{fig:ME_sig10mu1})
but then
returns to similar values as in the non-magnetic simulation once the magnetic field saturates to smaller values.
Plots of either axisymmetric or non-axisymmetric flow and kinetic energy spectra do not show visible differences
in the non-dynamo and dynamo cases. 

\begin{figure}
\centering
   \includegraphics[clip=true,width=0.4\textwidth]{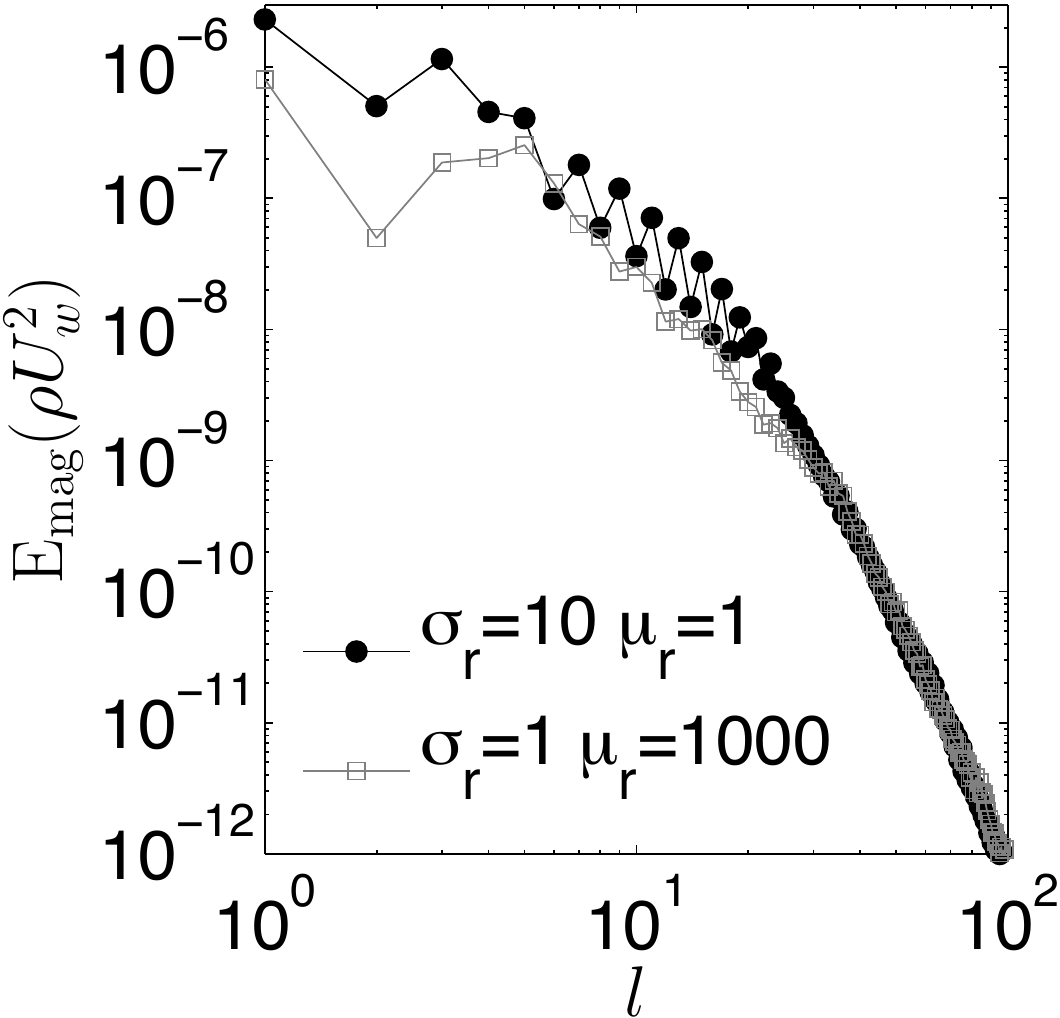}
 \caption{Poloidal magnetic energy spectra as a function of harmonic degree $l$ for \mbox{$(\sigma_r,\mu_r)=(10,1)$} (black circles) and 
 \mbox{$(\sigma_r,\mu_r)=(1,1000)$} (gray squares) and for \mbox{$\hat{h}=0.1$}. 
 The magnetic energy has been averaged in time and over the whole volume of the fluid.}
 \label{fig:spectre_ME}
\end{figure}

The magnetic field is mostly an axisymmetric toroidal field (about 80\% of the 
total magnetic energy). 
Figure~\ref{fig:spectre_ME} shows poloidal magnetic energy spectrum for the cases \mbox{($\hat{h},\sigma_r,\mu_r)=(0.1,10,1)$}
and \mbox{$(\hat{h},\sigma_r,\mu_r)=(0.1,1,1000)$}. The poloidal field is mostly dipolar ($l=1$).  
In all cases, the dipole is mainly axial and does not reverse polarity. 
We obtained very similar results for the magnetic field generated with a thin wall, \mbox{$\hat{h}=0.01$}.
Consequently, in the rest of the paper, we only describe the analysis of the results for the case \mbox{$\hat{h}=0.1$}.

The characteristics of the axisymmetric magnetic field are in good agreement with
the results described in RGC10.
We can also compare the topology of the self-sustained magnetic field in our work with 
other studies of von K\'arm\'an flows and the self-consistent generation of dynamos \citep{Bay07,Gis08b,Reu11}
although they have not addressed the role of magnetic boundary conditions.
These authors use spherical geometries and a volume forcing to mimic the role 
of disks rather than a boundary forcing. They find dynamos without the presence of a 
conducting wall between the outer sphere and the vacuum, but these dynamos only operate at larger magnetic Prandtl numbers 
than considered in this paper. 
\citet{Bay07} and \citet{Gis08b} also obtain an axisymmetric magnetic field (mainly an axial dipole for the poloidal part)
when the flow has a non-axisymmetric component and for magnetic Prandtl numbers of order unity.  
\citet{Reu11} show that when increasing the Reynolds numbers with fixed $\Pm$, 
the sustained magnetic field becomes small-scale. 
However they find that small Prandtl number calculations at fixed $\Rey$ yield a dipole dominated field.
In the study of \citeauthor{Reu11}, the highest Reynolds number, based on rms velocity, is \mbox{$\Rey^{\ast}=2367$}, for which
a small-scale dynamo is generated with a magnetic energy spectrum peaking around \mbox{$l=5$} for \mbox{$\Pm_f=0.25$}.
The rms Reynolds number is defined as \mbox{$\Rey^{\ast}=u^{\ast} \Rey$} where $u^{\ast}$
is the dimensionless rms velocity.
In our MHD calculations, the rms velocity is \mbox{$u^{\ast}\approx0.048$}, that is \mbox{$\Rey^{\ast}\approx2313$}.
All the dynamos we obtain at \mbox{$\Pm_f=0.01$} have fields dominated by the dipole component. 

\section{\label{sec:dynamo_mec} Analysis of the dynamo mechanism}

In order to elucidate the effect of the wall on the dynamo mechanism, 
it is necessary to describe in detail the generation of magnetic field in our simulations.
Since most of the magnetic energy resides in the axisymmetric components, and dynamos
are generally understood in a mean field framework \citep[\emph{e.g.} see the review of][]{Rob07},
we will focus on the production of the axisymmetric magnetic field, the largest scale of the system.
The role of a change in the wall magnetic properties
is described for two canonical cases: large wall conductivity, \mbox{$(\hat{h},\sigma_r,\mu_r)=(0.1,10,1)$}, hereafter called Case C,
and large wall permeability, \mbox{$(\hat{h},\sigma_r,\mu_r)=(0.1,1,1000)$}, hereafter called Case P. 
In Section~\ref{sec:omega_effect}, we show that the toroidal magnetic field
is induced by zonal velocity shear within the boundary 
layer adjacent to the outer wall. 
In both Cases C and P, the magnetic properties of the wall play a crucial role in 
allowing a strong toroidal field to develop in the shear layer.
In Section~\ref{sec:alpha}, we set out the ingredients that lead to the production of the mean emf, the source
of the poloidal field.
We find that only a limited range of azimuthal modes (between \mbox{$5\leq m \leq 14$}) contribute,
and we will propose an explanation for this observation based on
the properties of the flow. 
Finally, our explanations are verified in a set of numerical experiments in Section~\ref{sec:rolewall}.

\subsection{\label{sec:omega_effect} Generation of the axisymmetric toroidal field}
\begin{figure*}
\centering
   \subfigure[Case C: $(\sigma_r, \mu_r)= (10,1)$]{\label{fig:Baxi_sig10mu1}
   \includegraphics[clip=true,height=5cm]{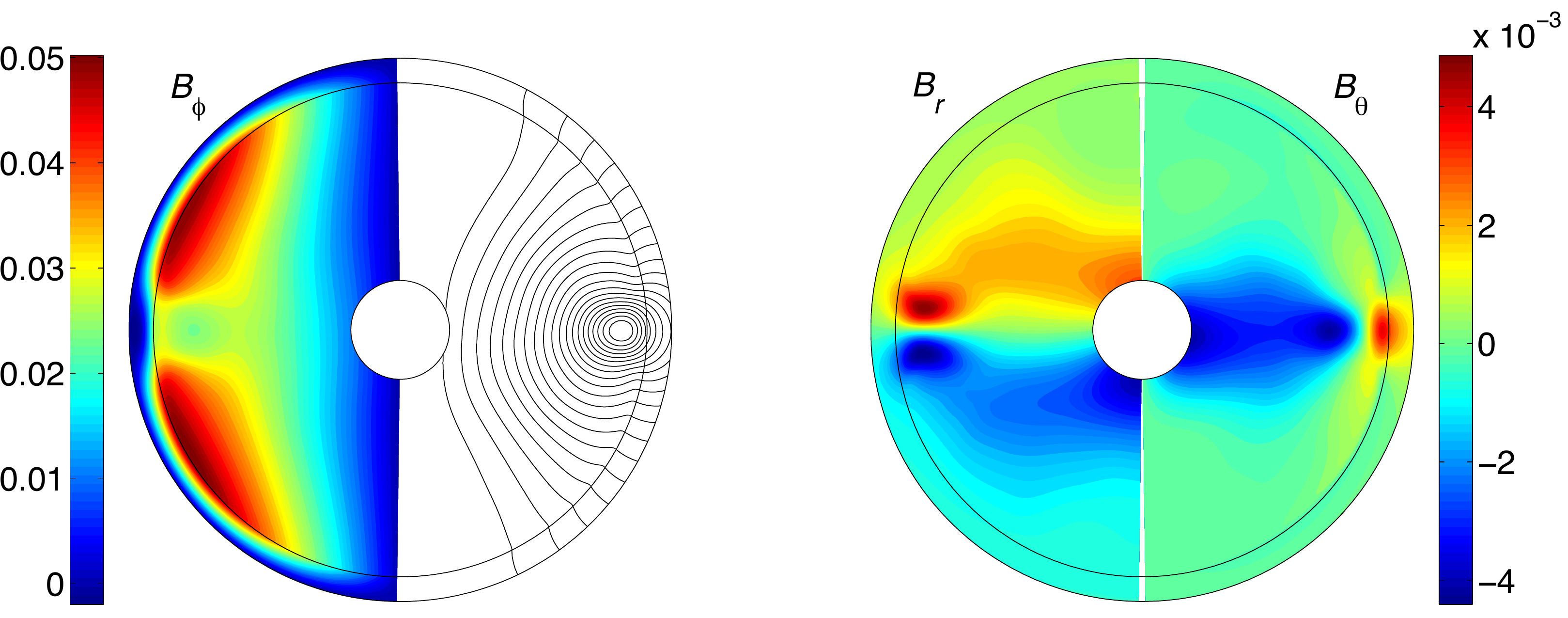}}  
   \subfigure[Case P: $(\sigma_r, \mu_r) = (1,1000)$]{\label{fig:Baxi_sig1mu1000}
   \includegraphics[clip=true,height=5cm]{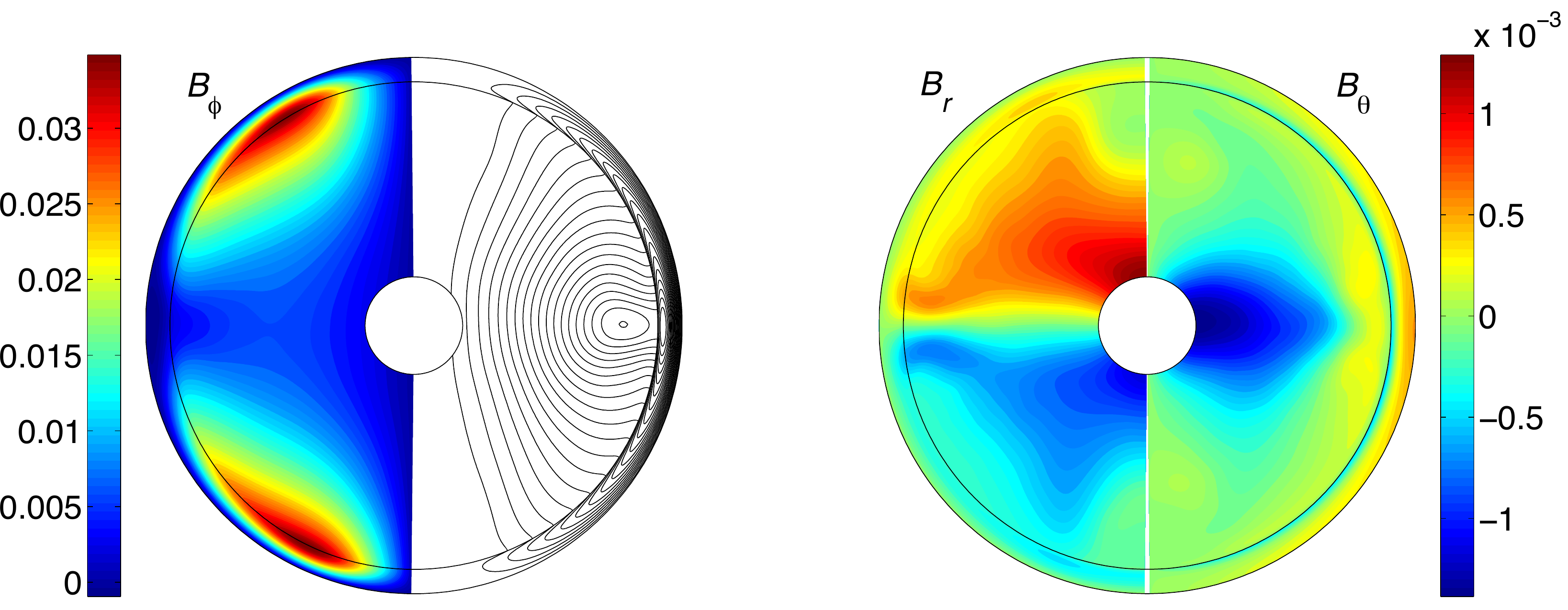}}  
\caption{Color. Axisymmetric magnetic field in a meridional plane (time-average over about \mbox{100$r_0/U_w$}). 
From left to right: Azimuthal component $\overline{B_{\phi}}$, poloidal magnetic field lines, 
radial component $\overline{B_r}$ and latitudinal component $\overline{B_{\theta}}$. 
The magnetic field is given in units of mbox{$\sqrt{\rho \mu_0} U_w$}. 
The outer shell represents the wall of thickness $\hat{h}=0.1$.
For the color plots of Case P, $\overline{B_{\phi}}$ in the wall is divided by 1000
and $\overline{B_{\theta}}$ by 10.}
\label{fig:Baxi}
\end{figure*}

The axisymmetric dynamo magnetic field components in Cases C and
P are plotted in Figure~\ref{fig:Baxi}.
In both cases, the axisymmetric azimuthal magnetic field, $\overline{B_{\phi}}$,
is generated in the narrow fluid shear layer next to the outer wall. 
The axisymmetric poloidal magnetic field intersects this shear layer, and experiences 
the strong radial gradient of $\overline{u_{\phi}}$.
The toroidal field is then created mostly by the so-called $\omega$ effect \citep[\emph{e.g.}][]{Rob07}, 
where $\overline{B_{\phi}}$ is induced by the action of the radial shear of $\overline{u_{\phi}}$ on
the axisymmetric radial magnetic field, $\overline{B_r}$.

\setcounter{subfigure}{0}

The toroidal field must vanish in the vacuum region outside of the wall, \emph{i.e.} \mbox{$\overline{B_{\phi}}(r_o+h)=0$}.
The wall therefore provides a buffering region between 
the fluid and the vacuum for $\overline{B_{\phi}}$; in the absence of the conducting outer wall,
the toroidal field would necessarily vanish at $r=r_o$, implying the presence of large radial gradients of $\overline{B_{\phi}}$
in the fluid shear layer and so strong ohmic dissipation there.
From Equations~(\ref{eq:Btan}) and~(\ref{eq:jtan}) we can deduce the following continuity equations for $\overline{B_{\phi}}$ 
and its radial gradients at the fluid-wall interface:
\begin{eqnarray}
	\left. \overline{B_{\phi}} \right|_w &=& \left. \mu_r \overline{B_{\phi}} \right|_f,
	\\
	\left. \frac{\partial r\overline{B_{\phi}}}{\partial r} \right|_w &=& \sigma_r \mu_r \left. \frac{\partial r\overline{B_{\phi}}}{\partial r} \right|_f ,
\end{eqnarray} 
where the discontinuity of \mbox{$\partial_r r\overline{B_{\phi}}$} is deduced from the discontinuity of the tangential electric currents, 
\mbox{$\overline{j_{\theta}}=-(1/r) \partial_r (r \overline{B_{\phi}}/\mu)$}.
High values of $\sigma_r$ or $\mu_r$ buffer $\overline{B_{\phi}}$ in the fluid in different ways, as follows.

\subsubsection{Effect of large conductivity: Case C}
\begin{figure*}
\centering
   \subfigure[$j_{P}$ for $\sigma_r=10$]{\label{fig:j_sig10mu1}
   \includegraphics[clip=true,width=0.45\textwidth]{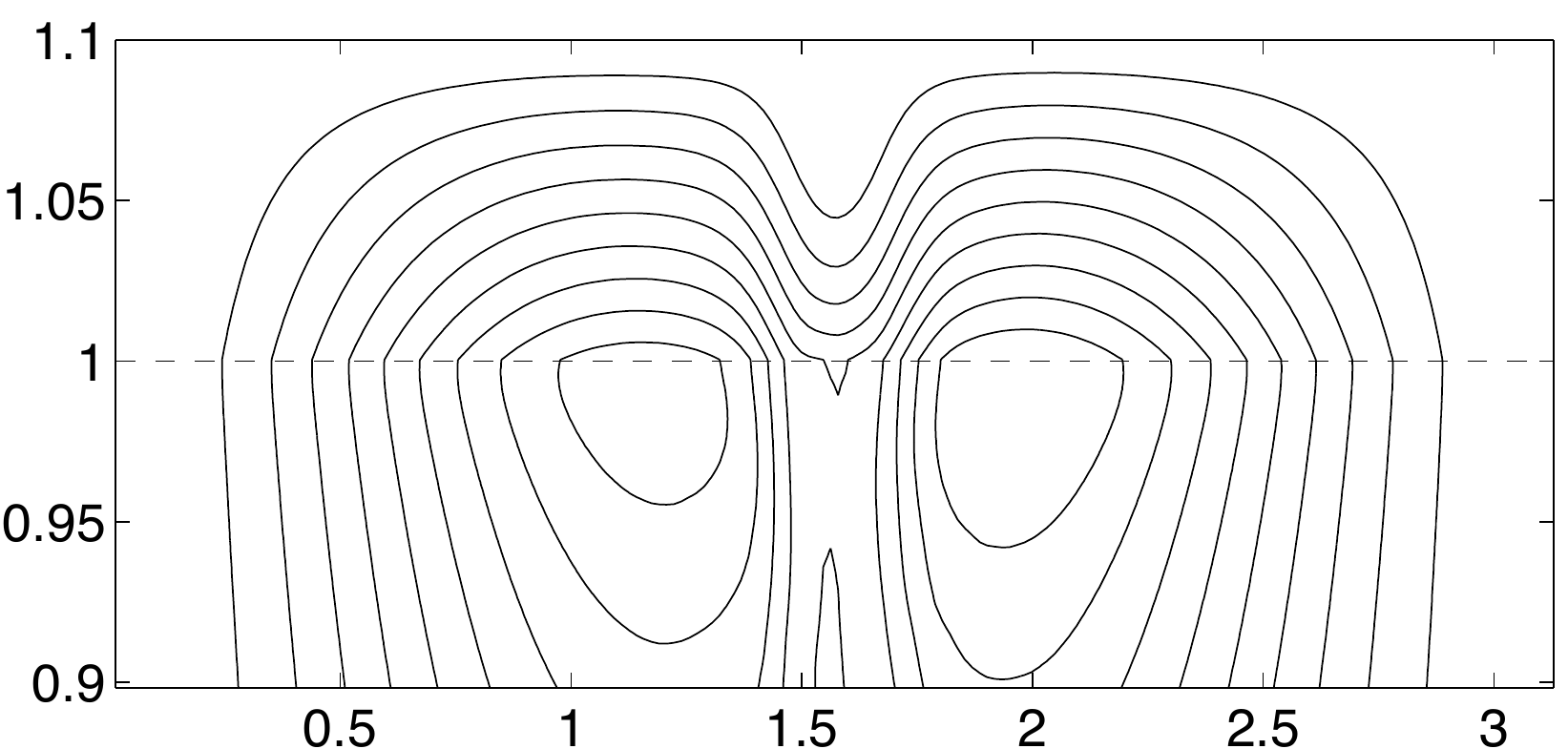}}
   \subfigure[$j_{P}$ for $\sigma_r=100$]{\label{fig:j_sig100mu1}
   \includegraphics[clip=true,width=0.45\textwidth]{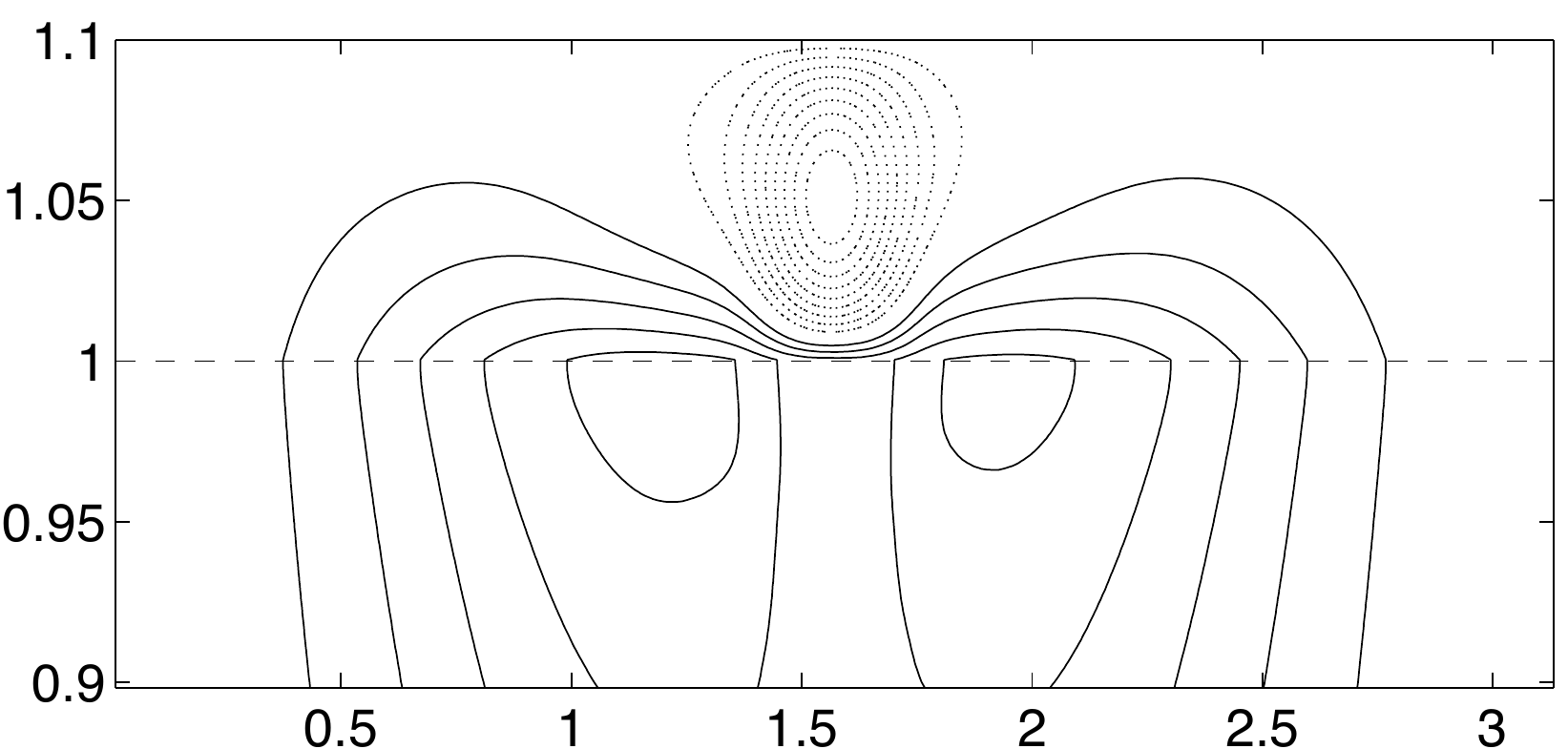}}   
   \subfigure[$\overline{B_{\phi}}(\theta=\pi/4,r)$]{\label{fig:Bpcut_sig}
   \includegraphics[clip=true,width=0.5\textwidth]{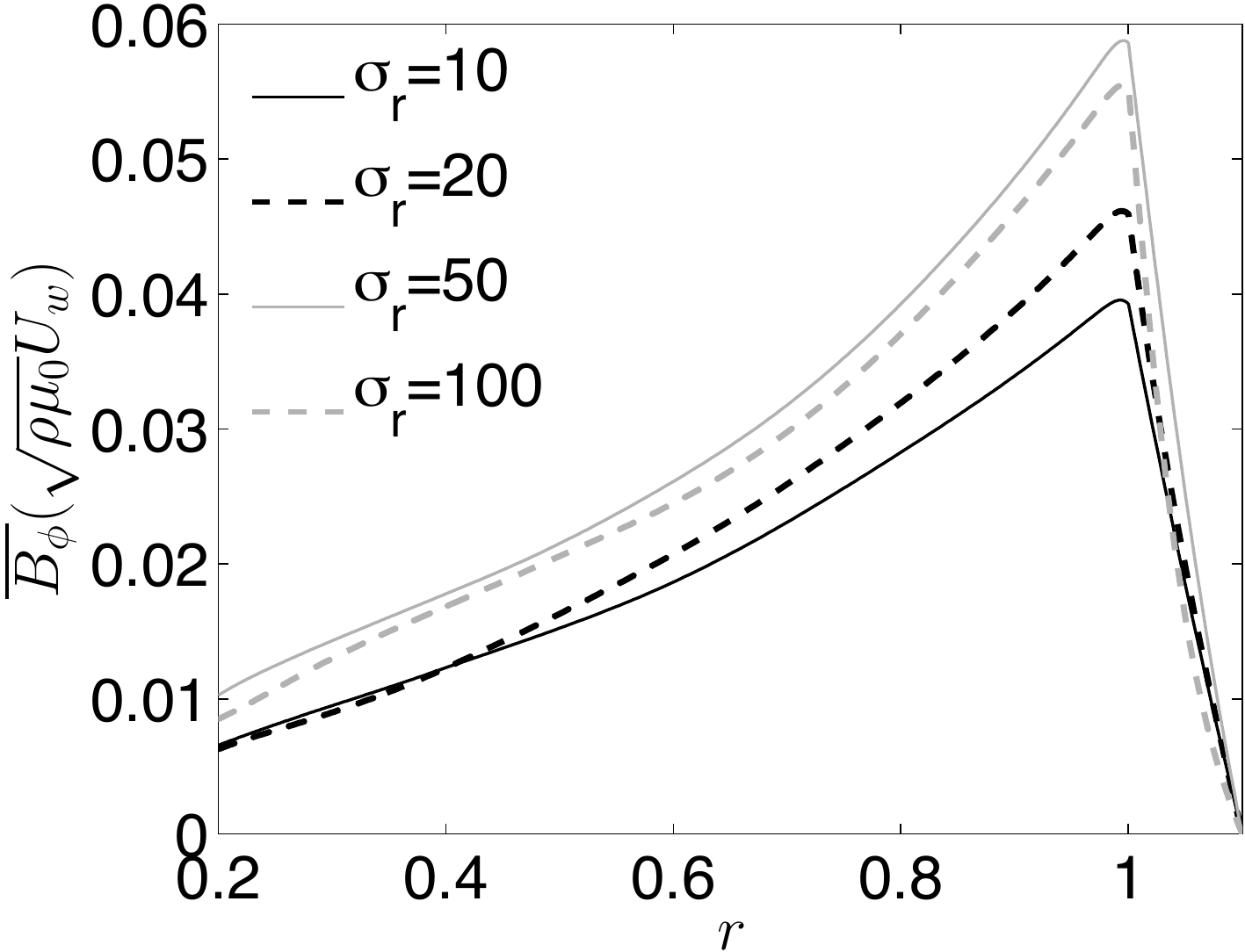}}
\caption{(a) and (b) Axisymmetric poloidal electric currents, $j_P$, in a close-up of the meridional plane for 
$\mu_r=1$. 
The vertical and horizontal axis corresponds to the radius and the colatitude respectively.
The solid (dotted) lines represent clockwise (counter-clockwise respectively) electric currents.
(c) \mbox{$\overline{B_{\phi}}(\theta=\pi/4,r)$} for \mbox{$\mu_r=1$} and varying $\sigma_r$.
The region $r>1$ corresponds to the wall.}
\end{figure*}

Case C has \mbox{$\mu_r=1$}, and so $\overline{B_{\phi}}$ is continuous, 
but the relatively high conductivity of the wall allows for large radial gradients of $\overline{B_{\phi}}$ in the wall
compared to the fluid. 
In this case, a large amplitude of $\overline{B_{\phi}}$ in the shear layer can match to the vacuum boundary condition
with weak radial gradients of $\overline{B_{\phi}}$ in the fluid near the wall.
High wall conductivity therefore leads to less limitation on the growth of $\overline{B_{\phi}}$ in the shear layer. 
Equivalently, large $\sigma_r$ enables the circulation of large latitudinal electric currents in the wall, which supports an 
axisymmetric azimuthal field of large amplitude in the fluid (Figure~\ref{fig:j_sig10mu1}).
 
These ideas are borne out in Figure~\ref{fig:Bpcut_sig}, which
shows the radial profile of the axisymmetric azimuthal field at the colatitude \mbox{$\theta=\pi/4$}
for increasing values of $\sigma_r$ and \mbox{$\mu_r=1$}.
The maximum of the azimuthal field in the fluid increases up to \mbox{$\sigma_r=50$} but 
starts to saturate for higher $\sigma_r$.
It appears that for \mbox{$\sigma_r \geq 50$}, the wall entirely shields 
the fluid from the vacuum boundary condition and the amplitude of the azimuthal
field, unrestrained by the vacuum boundary condition, 
only depends on the induction and its competition with magnetic diffusion in the shear layer.
For \mbox{$\sigma_r=10$} (Figure~\ref{fig:j_sig10mu1}), the electric currents fill the whole thickness of the wall.
For \mbox{$\sigma_r=100$} (Figure~\ref{fig:j_sig100mu1}), the currents appear more confined to the inner side of the wall at high latitudes.
This is explained by the discontinuity of the tangential currents at the interface (Equation~\ref{eq:jtan})
leading to stronger values of $\overline{j_{\theta}}$ in the inner part of the wall.
A counter-clockwise current loop is also present in the outer equatorial part of the wall.

\setcounter{subfigure}{0}

The wall thickness plays here a similar role to the wall conductivity: a thick wall makes 
the matching between a large amplitude of $\overline{B_{\phi}}$ in the shear layer and the vacuum possible.
The comparison of our calculations at \mbox{$\hat{h}=0.1$} and $\hat{h}=0.01$ in Figure~\ref{fig:dynamo_space} confirms 
the favorable role of a thick wall on dynamo action.
However, we note that for \mbox{$\hat{h}>0.1$} this favorable effect may well be lost due to the occurrence of a skin effect for oscillatory magnetic fields
as shown by \citet{Kai99}.

In summary, for a wall of small magnetic permeability, a high conductivity or a large thickness 
is necessary to obtain a large 
amplitude of the axisymmetric toroidal field due to the proximity between the shear layer and the vacuum.

\subsubsection{Effect of large permeability: Case P}
When the wall has a large magnetic permeability, 
the discontinuity of the tangential magnetic field component across the fluid-wall interface (Equation~\ref{eq:Btan})
yields weak values of $B_{\theta}$ in the fluid close to the wall.
This forces the poloidal magnetic field in the fluid to be mainly radial, effectively
strengthening $\overline{B_r}$ in the shear layer (Figure~\ref{fig:Baxi_sig1mu1000}).
By maintaining this radial magnetic field across the shear layer, 
especially at high latitudes where the radial gradients of $\overline{u_{\phi}}$ are large,  
the wall directly enhances the $\omega$ effect via the induction term 
\mbox{$\overline{B_r} \partial_r \overline{u_{\phi}}$}, leading to
the production of a strong azimuthal magnetic field in the fluid. 
Since $\overline{B_r}$ is stronger at high latitudes in Case P than in Case C, 
the maximum values of $\overline{B_{\phi}}$ in the shear layer are observed 
at higher latitudes in Case P than in Case C, as can be seen in Figure~\ref{fig:Baxi}.
For Case P, the discontinuity of $\overline{B_{\phi}}$ at the interface implies that
a large toroidal magnetic field is present in the wall.
Furthermore, the discontinuity of $\overline{B_{\theta}}$ provokes 
an abrupt change in the direction of the poloidal magnetic field lines in the wall which connect
in the wall rather than in the vacuum, confining the magnetic field inside the wall. 

The arguments presented in this section assume that only diffusive processes are acting in the wall,
and so the radial gradient of $\overline{B_{\phi}}$ is roughly linear in $r$ within the wall.
An induction process for $\overline{B_{\phi}}$, due to the coupling of $\overline{u_{\phi}}$ and the axisymmetric poloidal
magnetic field, also happens in the wall but we have verified that the induction terms are several orders 
of magnitude smaller than the diffusive terms in the wall for each case.

\subsection{\label{sec:alpha} Generation of the axisymmetric poloidal field}
We have seen that the axisymmetric toroidal magnetic field, $\overline{B_T}$ is 
mainly generated from an $\omega$ effect acting on
the axisymmetric poloidal magnetic field, $\overline{B_P}$ (in particular the axial dipole).
According to the Cowling anti-dynamo theorem \citep[\emph{e.g.}][]{Rob07}, 
$\overline{B_P}$ must be sustained
by the coupling of non-axisymmetric velocity and magnetic modes.
Indeed we find that the poloidal magnetic field is of strongest amplitude in the equatorial belt where the non-axisymmetric
motions are present (Figure~\ref{fig:Baxi}). 
The equations for the evolution of the components of the axisymmetric poloidal field, $\overline{B_r}$ and $\overline{B_{\theta}}$, in the fluid are
\begin{eqnarray}
	\pdt{\overline{B_r}} &=& 
		\frac{1}{r \sin \theta} \frac{\partial}{\partial \theta} 
		\left( \sin \theta \mathcal{E} \right)
		- \left[ \nabla \times \frac{1}{\sigma_f} \nabla \times \frac{\overline{\mathbf{B}}}{\mu_0} \right]_r ,
\\
	\pdt{\overline{B_{\theta}}} &=& 
		- \frac{1}{r} \frac{\partial}{\partial r} 
		\left( r \mathcal{E} \right)
		- \left[ \nabla \times \frac{1}{\sigma_f} \nabla \times \frac{\overline{\mathbf{B}}}{\mu_0} \right]_{\theta} ,
\end{eqnarray}
where the source of the axisymmetric poloidal field, 
the mean electromotive force (emf), is
\begin{eqnarray}
	\mathcal{E} = \sum\limits_{m=0}^{\textrm{Mmax}} \mathcal{E}^m \quad \textrm{with} \quad 
	\mathcal{E}^m = \overline{u_r^m B_{\theta}^m - u_{\theta}^m B_r^m}.
\end{eqnarray}
$\mathcal{E}^m$ is non-zero if there are correlations between the velocity and magnetic field 
modes of same azimuthal order $m$, denoted here as $\vect{u}^m$ and $\vect{B}^m$.

The production of the mean emf $\mathcal{E}$ is a very complex problem, and so 
we break down the problem into steps, which are 
summarized below and then described in detail after. 
Step 1 consists of the main observation, which we
then explain by going through Steps 2, 3 and 4.
\begin{itemize}
	\item[Step 1] We find that the main contribution to the emf comes from a limited number of azimuthal modes,
	specifically $5 \leq m \leq 14$, which we will call the ``dynamo'' modes. 
	These modes, together with the axisymmetric mode $m=0$, can each sustain the dynamo via the mechanism as explained below.
	\item[Step 2] The components $B_r^m$ and $B_{\theta}^m$
	are mainly produced by the distortion of the axisymmetric toroidal magnetic field, $\overline{B_T}$,
	by the velocity modes of same azimuthal order, $u_r^m$ and $u_{\theta}^m$. 
	Consequently, $\vect{B^m}$ modes are out of phase by $\pi/2$ in $\phi$ with $\vect{u^m}$ modes of same direction ($r$ or $\theta$).
	\item[Step 3] As a consequence of Step 2, the emf $\mathcal{E}^m$ produced by the mode $m$ is non zero if $u_r^m$ and $u_{\theta}^m$,
	are partly out of phase. The latitudinal and radial gradients in the zonal velocity lead to this required systematic phase shift 
	between $u_r^m$ and $u_{\theta}^m$ because they are mainly located at different radii. 
	\item[Step 4] Dynamo modes with a narrow range of $m$ are selected because the phase shift between $u_r^m$ and $u_{\theta}^m$ 
	is only significant for modes with a typical shearing timescale of the same order as their turnover timescale.
\end{itemize}
 
Cases C and P display very similar
features for their emf so the mechanisms of generation of $\overline{B_P}$ 
are likely the same. 
Therefore, in the following, we only analyze Case C. 
This suggests that the wall magnetic properties play only a minor role
in the generation of the axisymmetric poloidal field from non-axisymmetric modes. 

\subsubsection*{Step 1: Main contributions to the emf}

\begin{figure*}
\centering
    \subfigure[$\mathcal{E}^m$ in a meridional plane]{\label{fig:emf}
    \includegraphics[clip=true,width=\textwidth]{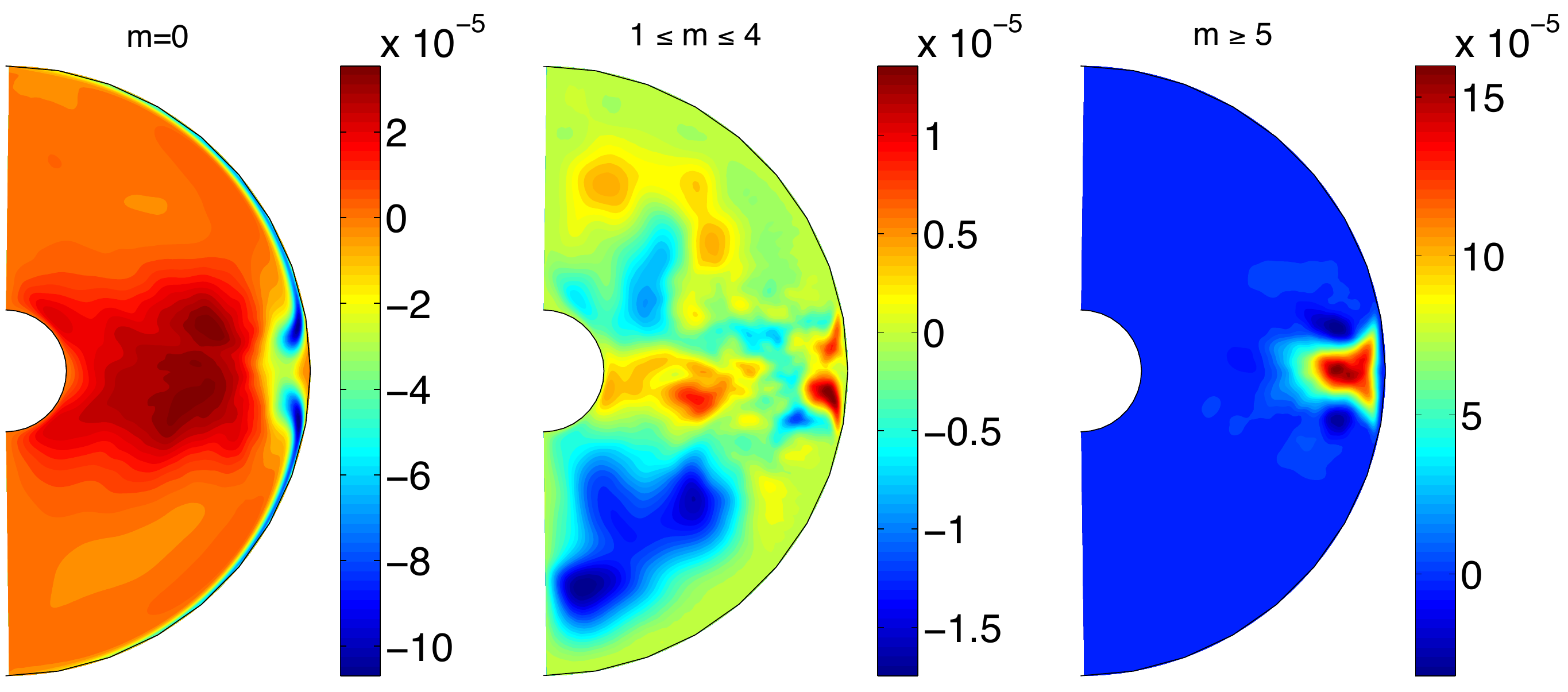}} 
    \subfigure[maximum of $\mathcal{E}^m$ from each mode]{\label{fig:graph_emf}
    \includegraphics[clip=true,height=5.cm]{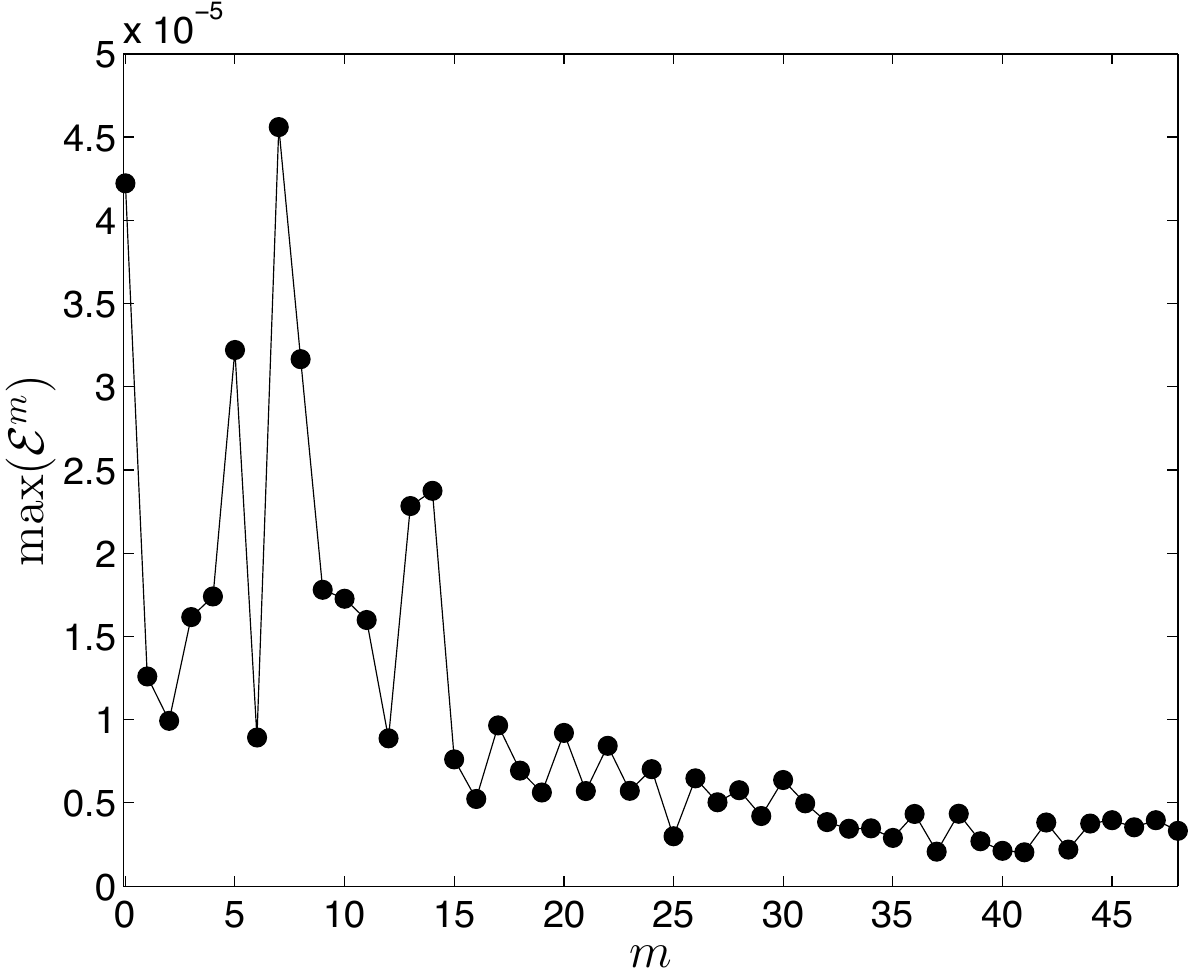}}
    \subfigure[magnetic energy spectrum]{\label{fig:spectre_mag_m}
    \raisebox{-0.2cm}{\includegraphics[clip=true,height=5.cm]{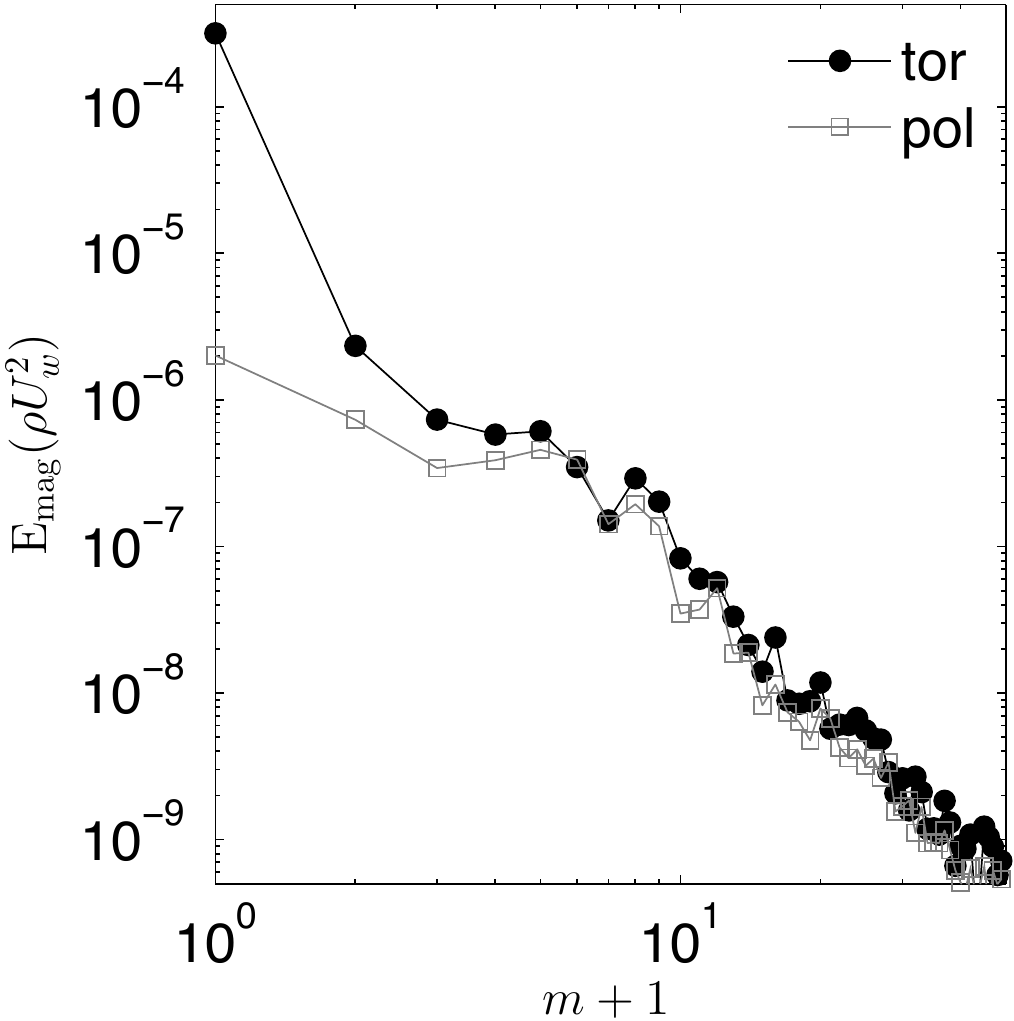}}}
\caption{Color. (a)-(b) Time-averaged mean emf $\mathcal{E}$ and (c) magnetic energy spectrum for 
azimuthal modes in the toroidal component (black circles) and in
the poloidal component (gray squares) for Case C.}
\end{figure*}

Figure~\ref{fig:emf} shows the time-averaged emf produced by the interactions of
\mbox{$m=0$} (axisymmetric) modes and other groups of the \mbox{$m>0$} (non-axisymmetric) modes. 
As expected from Cowling's theorem, the largest values of $\overline{B_r}$ and $\overline{B_{\theta}}$ (Figure~\ref{fig:Baxi_sig10mu1}) 
do not correlate
with the latitudinal and radial derivatives of $\mathcal{E}^{m=0}$, respectively.
Figure~\ref{fig:emf} also shows that
$\mathcal{E}^m$ from the \mbox{$m\ge 5$} modes is one order of magnitude larger than from the \mbox{$1 \le m \le 4$} modes,
and is produced in the equatorial belt
with its latitudinal and radial derivatives well correlated with $\overline{B_r}$ and $\overline{B_{\theta}}$ in Figure~\ref{fig:Baxi}.
Figure~\ref{fig:graph_emf} shows the maximum of $\mathcal{E}^m$ from each mode; clearly 
some modes produce a significantly larger emf than others, most notably \mbox{$m=5,7,8,13,14$}.
However, the kinetic energy of these modes is not noticeably different than the other modes according to
the kinetic energy spectrum, which displays a nearly flat slope for \mbox{$m\le8$} (Figure~\ref{fig:Spectrum_KE_Re48193}).
Furthermore the magnetic energy spectrum in $m$ (Figure~\ref{fig:spectre_mag_m}) also shows
that the azimuthal modes producing a large emf do not exhibit a significantly larger magnetic energy. 
The \mbox{$m\ge15$} modes produce weak contributions to the emf. 

To explain why the production of a temporally- and spatially-coherent 
emf depends on the azimuthal order of the modes, we
first studied whether a given $m$ mode can sustain the axial dipole on its own.
To do so, we ran a simulation identical to Case C, except that 
only modes \mbox{$m=0, 5, 10, 15,...$} are calculated.
We obtained a dynamo with an axisymmetric magnetic field of very similar characteristics 
to that of Case C. This simulation has the same toroidal field morphology, an axial dipole with strong values 
of $\overline{B_r}$ in the equatorial belt, and about half the amplitude of the axisymmetric field produced in Case C.
We further ran a similar numerical simulation but this time calculating only modes \mbox{$m=0, 20, 40, ...$}.
This case fails to produce a dynamo.
We conclude that for selected $m$ the production of $\mathcal{E}^m$ depends only on
the interaction of a mode with itself
and with the axisymmetric magnetic field. 
Hereafter these selected modes are called the dynamo modes, and we
consider their contributions to the emf individually.
We now need to understand how the dynamo modes organize to produce a 
coherent emf and what physical mechanism determines which azimuthal orders are dynamo modes.

\subsubsection*{Step 2: Generation of the non-axisymmetric magnetic field}

\begin{figure*}
\centering
     \subfigure[radial components, $u_r^{m=5}$ and $B_r^{m=5}$]{
     \includegraphics[clip=true,width=\textwidth]{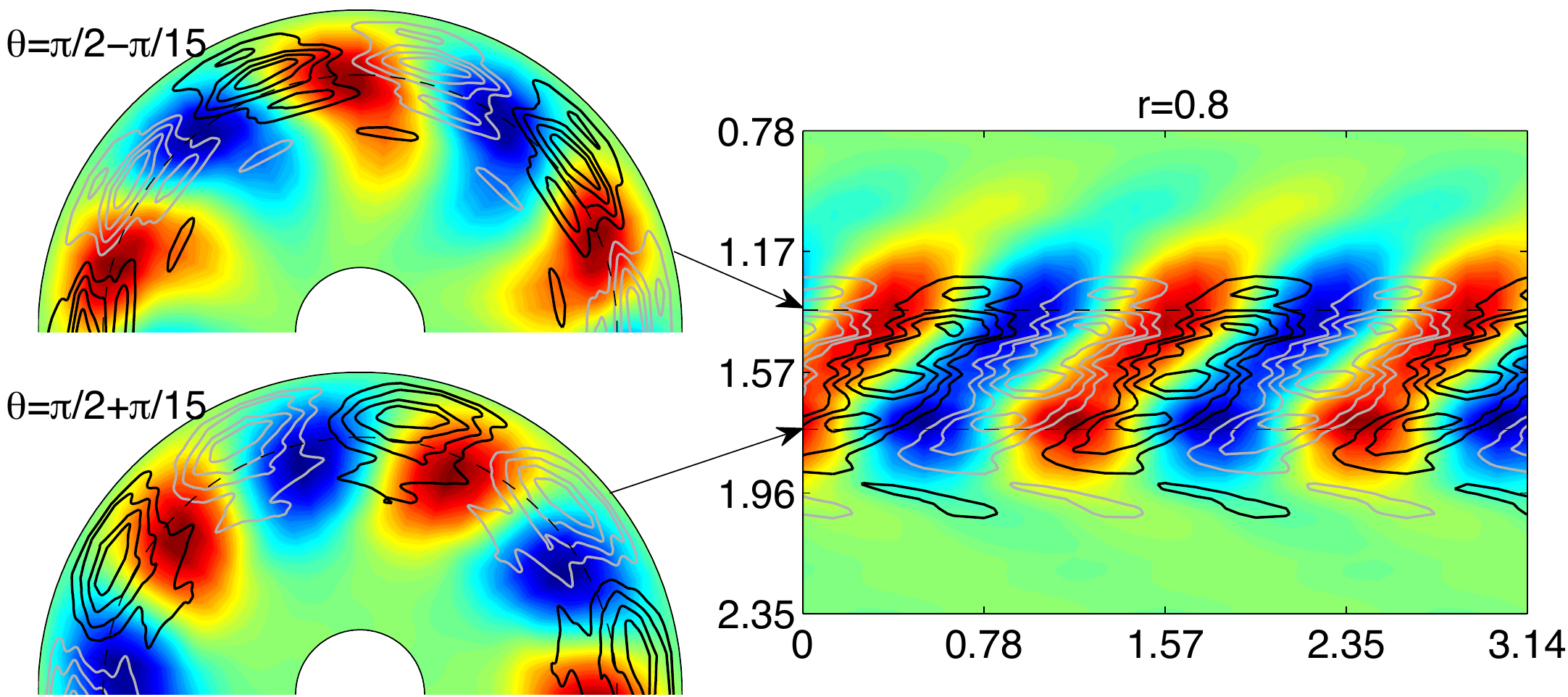}}
     \subfigure[latitudinal components, $u_{\theta}^{m=5}$ and $B_{\theta}^{m=5}$]{\label{fig:btut}
     \includegraphics[clip=true,width=\textwidth]{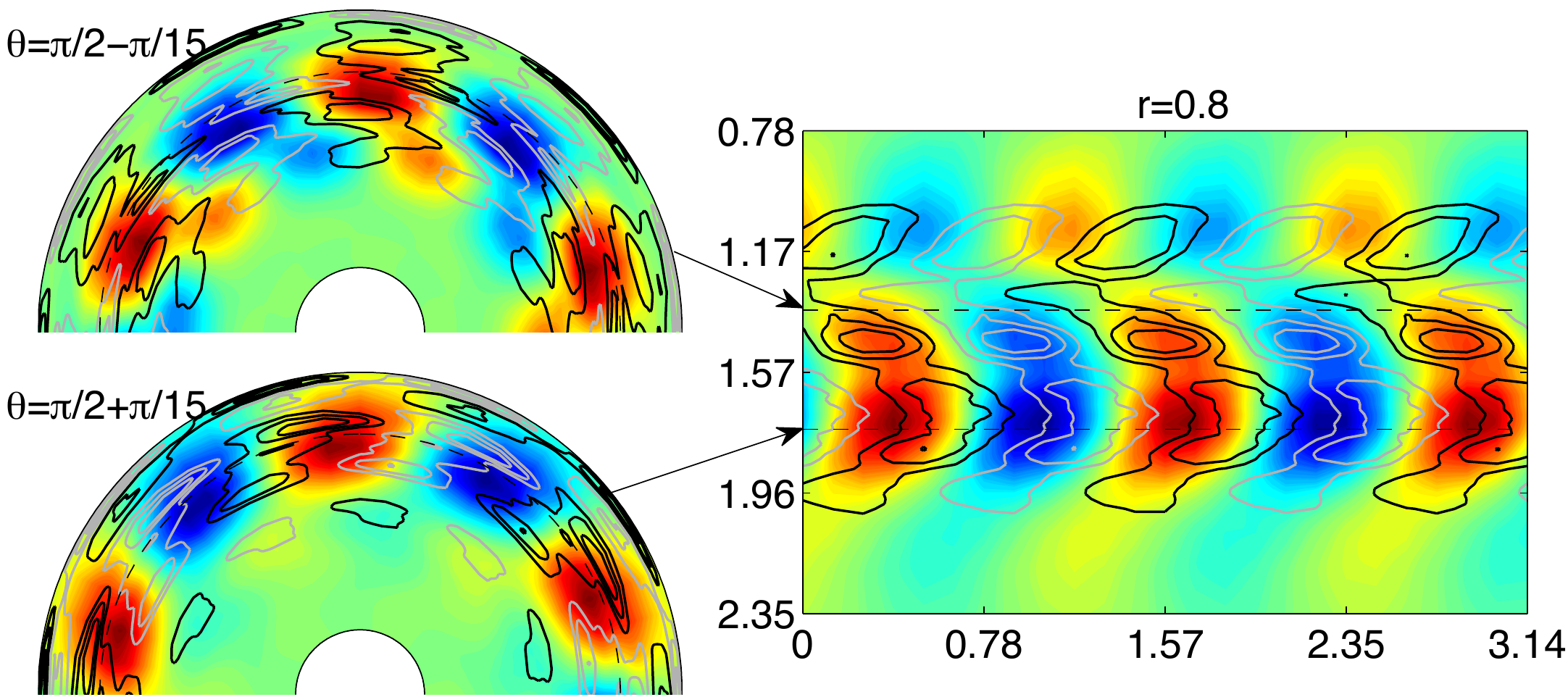}}
\caption{Color. Snapshot of the radial (a) and latitudinal (b) components of the $m=5$ magnetic (color) and velocity (lines) mode
(Case C).
Left: \mbox{$(r,\phi)$} planes above and below the equatorial plane at colatitudes \mbox{$\theta=\pi/2 \pm \pi/15$}.
Right: \mbox{$(\phi,\theta)$} plane at radius $r=0.8$ with the colatitude on the vertical axis
(range limited to \mbox{$[\pi/4,3\pi/4]$}) and the longitude on the horizontal axis (range limited to \mbox{$[0,\pi]$}).
For the magnetic field in color: red: positive and blue: negative.
For the velocity: black line: positive and gray line: negative.}
\label{fig:brur_btut}
\end{figure*}
\setcounter{subfigure}{0}

A magnetic mode $m$ is partly generated by the interaction of the axisymmetric magnetic field
with the $m$ velocity mode. Since \mbox{$\overline{B_{\phi}} \gg \overline{B_r}, \overline{B_{\theta}}$}
as observed in Figures~\ref{fig:ME} and~\ref{fig:Baxi},
the evolution rates of the non-axisymmetric field produced by this interaction, $B_r^m$ and $B_{\theta}^m$, are
given by,
\begin{eqnarray}
	\pdt{B_r^m} & \sim & \frac{\overline{B_{\phi}}}{r \sin \theta} \frac{\partial}{\partial \phi} u_r^m 
	+ \left[ \nabla^2 \vect{B}^m \right]_r,
	\label{eq:ind_Brm}
	\\
	\pdt{B_{\theta}^m} & \sim & \frac{\overline{B_{\phi}}}{r \sin \theta} \frac{\partial}{\partial \phi} u_{\theta}^m 
	+ \left[ \nabla^2 \vect{B}^m \right]_{\theta}.
	\label{eq:ind_Btm}
\end{eqnarray} 
The induction term on the right-hand side corresponds to the distortion of the axisymmetric 
toroidal magnetic field lines by the non-axisymmetric velocity modes. 
To assess if this interaction is the main source of the non-axisymmetric magnetic modes,
we compared
the induction terms (including all interactions) of $B_r^m$ and $B_{\theta}^m$ with the induction terms
in Equations~(\ref{eq:ind_Brm}) and (\ref{eq:ind_Btm}), respectively, for \mbox{$m=5$}. 
We found that the induction terms in Equations~(\ref{eq:ind_Brm}) and (\ref{eq:ind_Btm})
contribute to more than $70\%$ of the amplitude of the induction from all terms.
In Equation~(\ref{eq:ind_Brm}), the balance between the induction term and
the evolution rate of $B_r^m$ and/or with the diffusion term
implies that $B_r^m$ and $u_r^m$ should be out of phase by $\pi/2$ in $\phi$.
Similar arguments based on Equation~(\ref{eq:ind_Btm}) yield
that $B_{\theta}^m$ and $u_{\theta}^m$ must be phase shifted by $\pi/2$ in $\phi$.
Figure~\ref{fig:brur_btut} shows snapshots of the radial and latitudinal components of the velocity and
magnetic field for the \mbox{$m=5$} mode in different planes. We indeed observe a phase shift of about $\pi/2$ in $\phi$
between velocity and magnetic modes. 
Any snapshot of the \mbox{$m \ge 2$} modes shows similar results.

\subsubsection*{Step 3: Spatial distribution of the non-axisymmetric velocity}
The azimuthal average of \mbox{$u_r^m B_{\theta}^m$} in $\mathcal{E}$ is non zero 
if $u_r^m$ and $B_{\theta}^m$ are partly in phase in azimuth. 
Similarly a non zero azimuthal average of \mbox{$u_{\theta}^m B_r^m$} requires that $u_{\theta}^m$ and $B_r^m$ are partly in phase.
According to our previous argument, $u_r^m$ and $B_r^m$
are out of phase by $\pi/2$, and similarly for $u_{\theta}^m$ and $B_{\theta}^m$.
Hence $u_r^m$ must be at least partly out of phase with $u_{\theta}^m$ to obtain a non-zero emf.
The shearing caused by latitudinal and radial gradients of the zonal velocity
could create such a systematic phase shift.
This shearing of the non-axisymmetric velocity structures is visible in
Figure~\ref{fig:brur_btut} where the velocity components appear more noticeably slanted in
\mbox{$(\phi, \theta)$} planes than in \mbox{$(r, \phi)$} planes. 
In snapshot figures, the velocity modes appear to be sometimes torn apart in \mbox{$(\phi, \theta)$} planes 
as can be observed for $u_{\theta}^m$ in Figure~\ref{fig:btut}.
A systematic phase shift between $u_r^m$ and $u_{\theta}^m$
could be created when their strongest values have slightly different spatial locations
and therefore experience a different shear by $\overline{u_{\phi}}$.

\begin{figure*}
\centering
   \subfigure[$\overline{u_{\phi}}$]{
   \raisebox{0.35cm}{\includegraphics[clip=true,height=5.1cm]{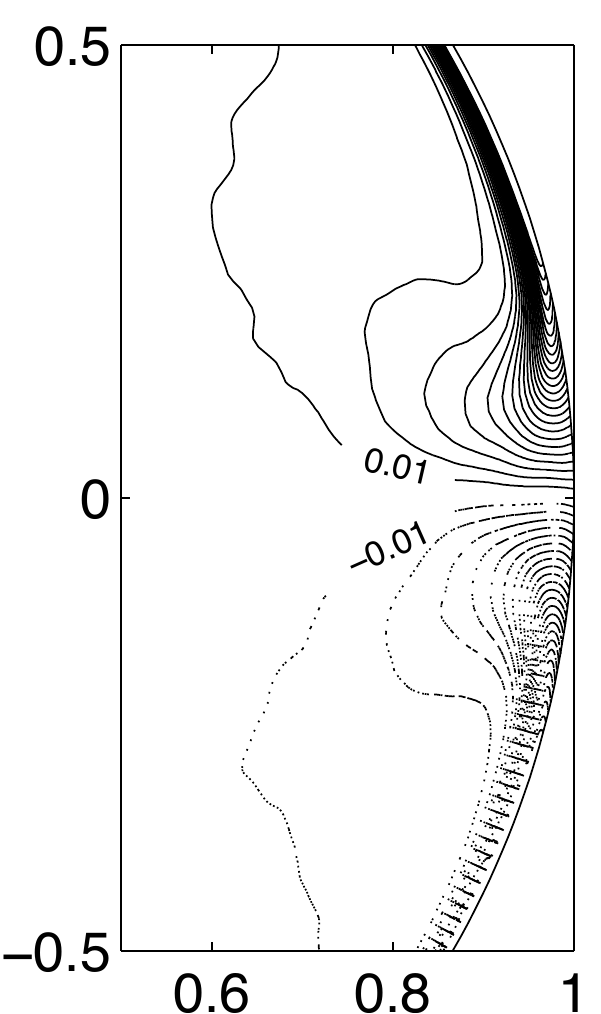}}}
   \subfigure[$|u_r|^{m>0}$]{
   \includegraphics[clip=true,height=5.6cm]{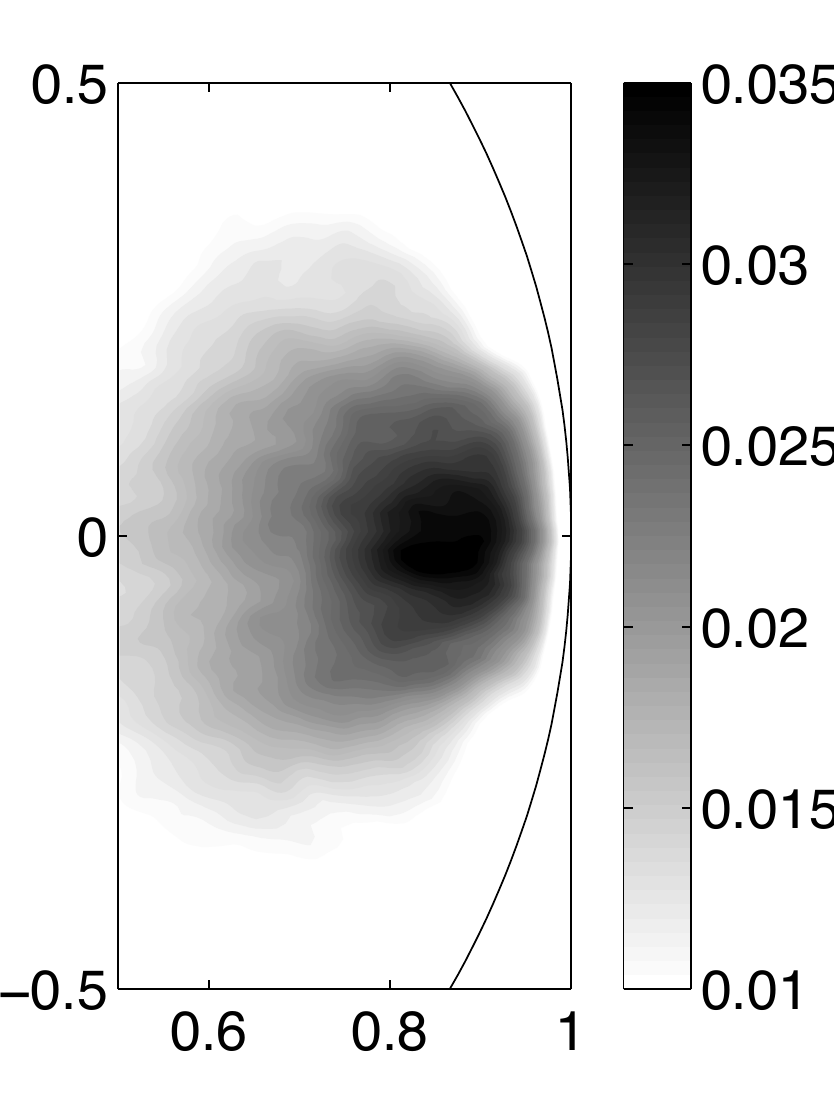}}
    \subfigure[$|u_{\theta}|^{m>0}$]{
   \includegraphics[clip=true,height=5.6cm]{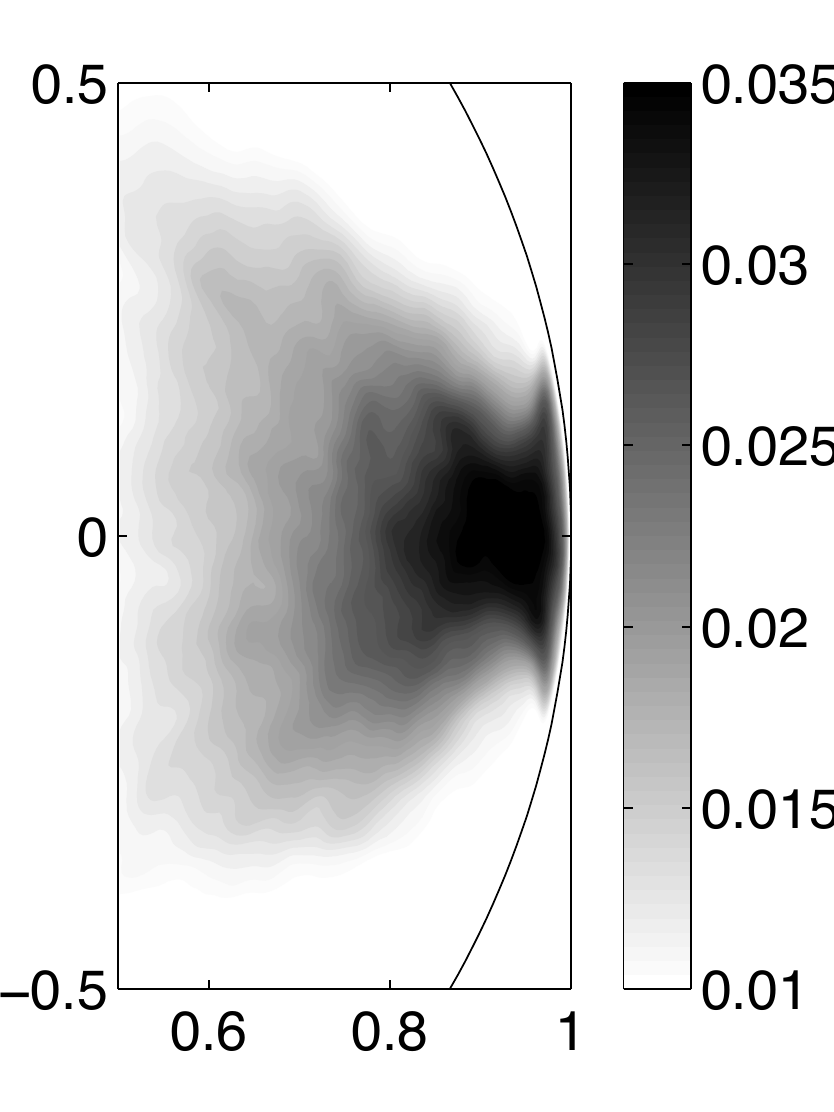}}  
\caption{(a) Zonal velocity, (b) radial and (c) latitudinal components of the non-axisymmetric rms velocity
(\mbox{$\sqrt{(|\vect{u}| - |\overline{\vect{u}}|)^2}$}) in a close-up of the meridional plane.
The contour interval for the zonal velocity is $0.01$; solid lines: positive and dashed lines: negative.}
\label{fig:amp_na}
\end{figure*}
\setcounter{subfigure}{0}

Figure~\ref{fig:amp_na} shows the non-axisymmetric rms velocity components in a meridional plane.
The strongest latitudinal velocity is located closer to the outer sphere than the strongest radial velocity, presumably
due to the presence of the impenetrable wall, which forces radial motions to decelerate at the wall and to recirculate tangentially.
Since latitudinal gradients of the zonal velocity are larger than radial gradients in the equatorial region, 
and the non-axisymmetric
structures are more visibly slanted in a \mbox{$(\phi,\theta)$} plane compared to a 
\mbox{$(r,\phi)$} plane (see Figure~\ref{fig:brur_btut}), 
the shearing caused by latitudinal gradient of $\overline{u_{\phi}}$ is predominant in the equatorial belt.
Since $u_{\theta}^m$ is located closer to the wall than $u_r^m$,
$u_{\theta}^m$ experiences a larger latitudinal shearing than $u_r^m$.

\subsubsection*{Step 4: Selection of the dynamo modes}

\begin{figure*}
\centering
\includegraphics[clip=true,width=0.9\textwidth]{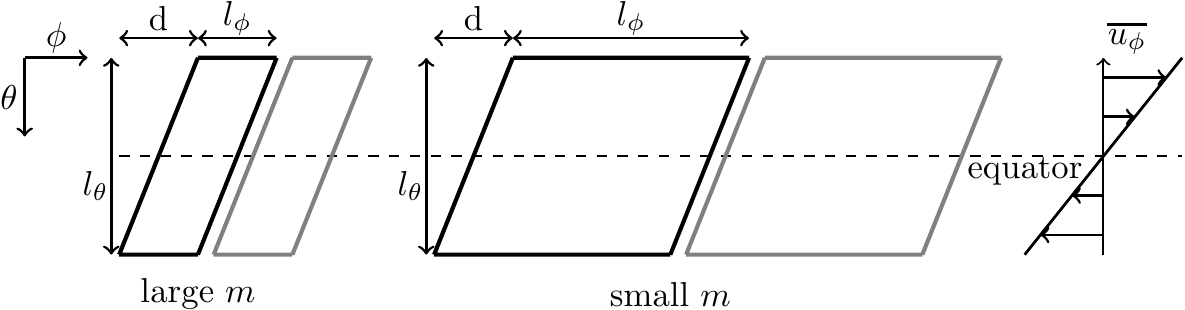}
\caption{Schematic representation of latitudinal velocity contours (black: positive and gray: negative)
in a plane \mbox{$(\phi,\theta)$} for two modes of different azimuthal order $m$.}
\label{fig:schema_shear}
\end{figure*}

For the moment, suppose that $u_r^m$ experiences no latitudinal shearing.   
To obtain a noticeable latitudinal shearing of $u_{\theta}^m$, the typical shearing timescale has to be comparable to
or shorter than the turnover timescale of the structure. This argument based on the dynamical timescales explains
why certain azimuthal modes are preferred.
To illustrate how the shearing affects differently structures depending on their azimuthal and latitudinal extents, 
Figure~\ref{fig:schema_shear} represents schematically contours of $u_{\theta}^m$ 
in a plane \mbox{$(\phi, \theta)$} at a given radius $R$ for two different azimuthal modes.
The ``shearing" distance, $d$, depends on the latitudinal gradient of the angular velocity, \mbox{$\Omega=\overline{u_{\phi}}/r \sin \theta$},
integrated over the latitudinal extent of the structure $l_{\theta}$ during a time $\Delta t$:
\begin{eqnarray}
	d=\frac{\partial \Omega}{\partial \theta} l_{\theta} \Delta t,
\end{eqnarray}
where $\Delta t$ is the typical turnover timescale:
\begin{eqnarray}
	\Delta t = \frac{l_{\theta}}{|u_{\theta}^m|}.
\end{eqnarray}
The shearing of the structure of azimuthal width \mbox{$l_{\phi} = 2\pi R/2m$} 
is significant if \mbox{$l_{\phi} \le d$}, that is, if
\begin{eqnarray}
	m \ge m_c = \frac{\pi R |u_{\theta}^m|}{\partial_{\theta} \Omega l_{\theta}^2}.
	\label{eq:mc}
\end{eqnarray} 
Note that, in general, we expect the latitudinal extent $l_{\theta}$ and the non-axisymmetric velocity amplitude 
$|u_{\theta}^m|$ to depend on $m$, so the formula~(\ref{eq:mc}) is non-linear in $m$.
Still supposing that $u_r^m$ is not sheared, the mode $m=m_c$ displays a phase shift of $\pi/2$ in $\phi$
between $u_r^m$ and $u_{\theta}^m$ in the equatorial plane. 
Modes with small $m$ experience a smaller relative deformation than larger $m$ modes for the same
shearing rate as pictured in Figure~\ref{fig:schema_shear}.
For modes with very large $m$, we may expect that the deformation undergone by the structures (\mbox{$d \gg l_{\phi}$}
since $l_{\phi}$ is small)
leads to an incoherent phase shift on average.

The simplified picture described here predicts  
the existence of a critical mode and possibly a critical range of modes (\mbox{$l_{\phi}\approx \mathcal{O}(d)$}), for which a systematic phase
shift between $u_r^m$ and $u_{\theta}^m$ is produced due 
to the equatorial anti-symmetry of the zonal flow. This argument
provides a consistent explanation for the observations made in Figures~\ref{fig:emf} and \ref{fig:graph_emf}:
(i) a coherent emf is produced only for \mbox{$m\ge 5$}, (ii) only some modes of selected azimuthal symmetry produce 
a large emf, and (iii) \mbox{$m\ge 15$} modes generate only a weak emf.

\begin{figure}
\centering
   \includegraphics[clip=true,width=0.45\textwidth]{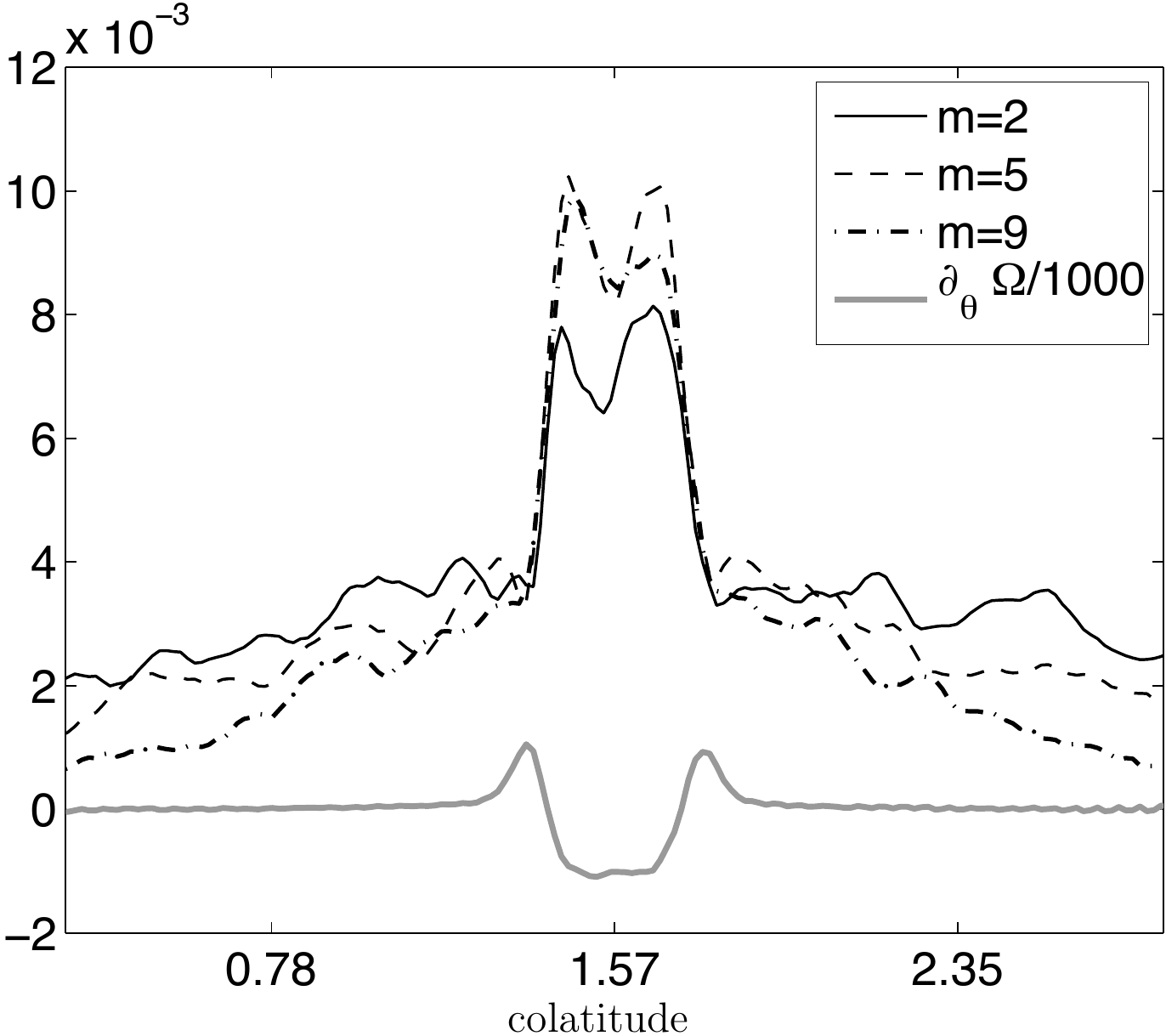}
\caption{Latitudinal profile of the rms latitudinal velocity at radius $r=0.96$ 
for different azimuthal modes $m$ (time and azimuthal average). 
The latitudinal gradient of angular velocity $\Omega$ at radius $r=0.96$ is plotted in gray (divided by 1000).} 
\label{fig:ut_lat_ext}
\end{figure}

To evaluate $m_c$ from the numerical simulations, 
we plot in Figure~\ref{fig:ut_lat_ext} 
the latitudinal profile of the rms $|u_{\theta}^m|$ for the modes $m=2$, $m=5$ and $m=9$
at a given radius and time- and $\phi$-averaged.
First, the amplitude of the velocity at this radius is comparable for the three modes.
Second, the latitudinal extent of the structures varies only weakly with $m$. 
The peaks of the profiles have a width corresponding to the region of 
maximum latitudinal gradient of the angular velocity $\Omega$, of latitudinal extent about $0.2r_o$. 
For this azimuthal order range ($m\le 9$),
the latitudinal extent, $l_{\theta}$, may be limited to the region of largest latitudinal gradient 
of $\Omega$, readily explaining why different azimuthal modes have similar latitudinal extents.
Using the values obtained in the numerical simulations,
\mbox{$|u_{\theta}^m| \approx 0.01$}, \mbox{$R\approx1$}, \mbox{$|\partial_{\theta} \Omega| \approx 1$}
and \mbox{$l_{\theta}\approx0.2$} (given in non-dimensional units), we obtain \mbox{$m_c \approx 1$} from the formula~(\ref{eq:mc}). 
However, in the emf plots of Figure~\ref{fig:emf}, we found that \mbox{$m\ge5$} is necessary to obtain a coherent emf.
This apparent discrepancy is probably explained by the neglect of the latitudinal shear of $u_r^m$ so far:
$u_r^m$ also undergoes a latitudinal shear, but less that $u_{\theta}^m$ since \mbox{$\partial_{\theta}\Omega$}
is weaker at smaller radius where $u_r^m$ is strongest (Figure~\ref{fig:amp_na}).
Consequently, it is not surprising that a larger deformation of the velocity structure than \mbox{$d\approx l_{\phi}$} is required 
to obtain a significant phase shift between radial and latitudinal velocity, leading to a larger $m_c$ in the numerical
simulations than given by our simplified 
picture summarized in the formula~(\ref{eq:mc}).

In the arguments presented here, we have emphasized the importance of the radial segregation between 
the maximum values of $u_r^m$ and $u_{\theta}^m$.
$B_r^m$ is induced locally at same radius as $u_r^m$, and similarly for $B_{\theta}^m$ and $u_{\theta}^m$.
However the cross product $\mathcal{E}$ requires that $B_r^m$ exists in the same region as $u_{\theta}^m$, and similarly
for $B_{\theta}^m$ and $u_r^m$. 
Magnetic diffusion, which is at least as important as magnetic induction 
at scales $m \geq 5$ (for which \mbox{$\Rm_l \lesssim1$}, see Section~\ref{sec:hydro})
alleviates this problem by coupling regions of large radial gradients of $B_r^m$ and $B_{\theta}^m$. 
The large-scale dynamo described here may therefore fail for larger magnetic Prandtl numbers 
(\emph{i.e.} smaller magnetic diffusivities) than considered here.

\begin{figure}
\centering
   \includegraphics[clip=true,width=0.45\textwidth]{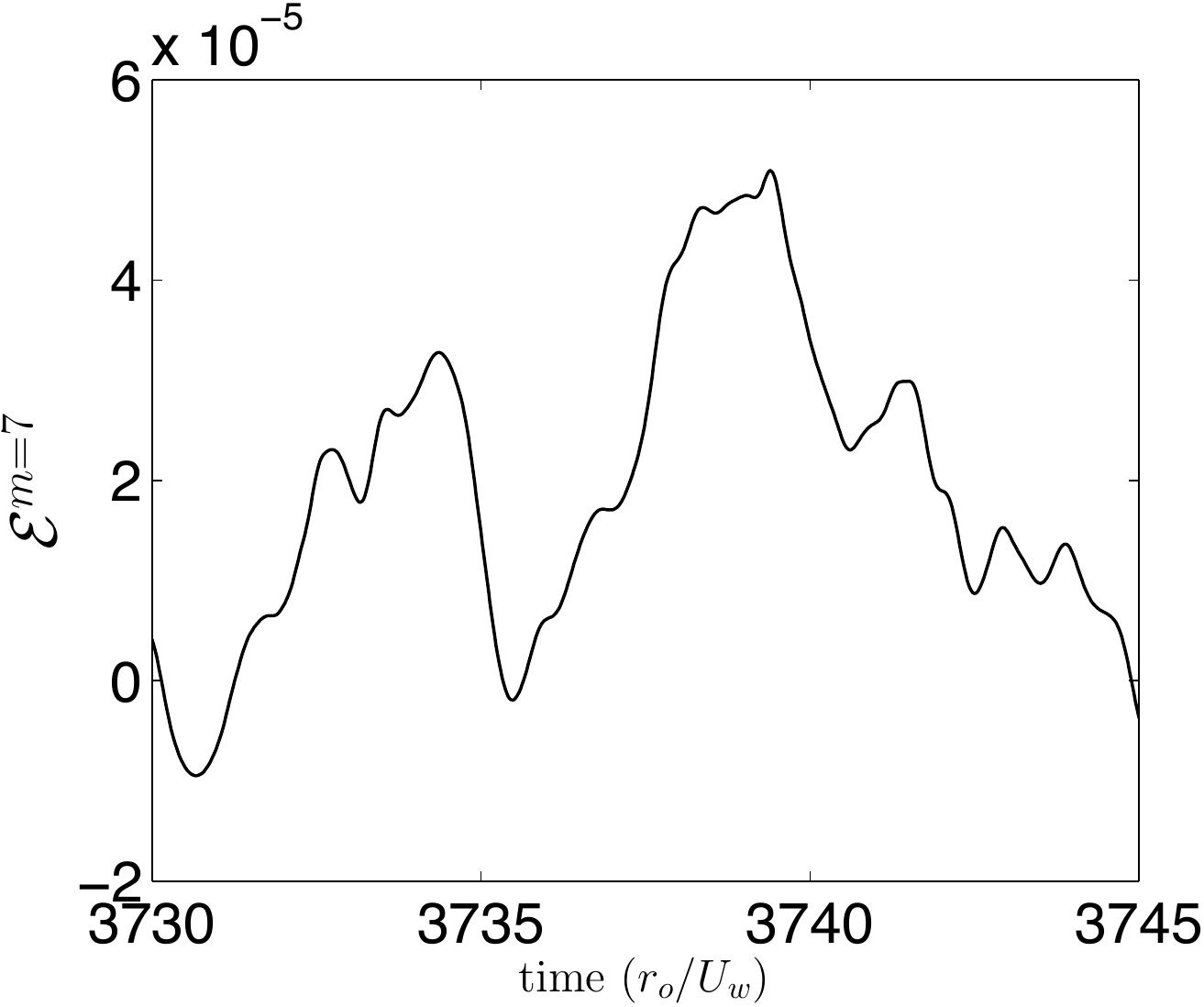}
\caption{Time series of $\mathcal{E}^m(r=0.85,\theta=\pi/2)$ produced by the $m=7$ mode.} 
\label{fig:emf_m7_t}
\end{figure}

Finally, we have not considered so far the fluctuating nature of the non-axisymmetric flow, which may 
lead to episodic losses of the phase shift between components of the velocity.
Figure~\ref{fig:emf_m7_t} shows a time series of \mbox{$\mathcal{E}^m(r=0.85,\theta=\pi/2)$} produced by the \mbox{$m=7$} mode.
$\mathcal{E}^m$ periodically goes to zero, but it is non zero on average.  
The contributions of the dynamo modes at different $m$ ensure that the total emf is always 
large enough to sustain the axial dipole.

In summary, we find that the axisymmetric poloidal field is generated by
a few non-axisymmetric modes. 
The latitudinal deformation of these modes, a consequence of the equatorial anti-symmetry of the zonal flow,
is crucial for the production of a coherent emf.
The selection of the non-axisymmetric dynamo modes depends on the boundary forcing (measured by the
Reynolds number) since this determines the latitudinal gradient of the zonal flow and the amplitude of 
the non-axisymmetric velocity (see Figure~\ref{fig:Ekin_forcing}).
All simulations here were performed at the same forcing but the influence of the Reynolds
number on the dynamo modes could be tested at a later date.

\subsection{\label{sec:rolewall}Effects of the wall magnetic properties on the non-axisymmetric modes}
In Section~\ref{sec:omega_effect}, we have demonstrated the importance of the wall
magnetic properties for the amplification of the axisymmetric toroidal field, both
by a large conductivity, which shields the shear layer from the vacuum, or by large permeability,
which directly enhances the $\omega$-effect by forcing the field to be normal in the shear layer.
We emphasize that the induction of the axisymmetric toroidal field clearly happens in the fluid shear layer. 
In Section~\ref{sec:alpha}, we have described the flow properties leading to the generation 
of an axisymmetric poloidal magnetic field without invoking the role of the wall magnetic properties.
This yields the tentative conclusion that the wall is necessary to the dynamo only to obtain a large amplitude 
axisymmetric toroidal field. This axisymmetric toroidal field then feeds the non-axisymmetric magnetic components responsible for the mean emf,
but the rest of the dynamo mechanism is ultimately a result of the flow properties.

\begin{table}
\begin{center}
\begin{tabular}{ | c | c | c | c | c | c |}
\hline
Case & $\sigma_r^{m=0}$ & $\sigma_r^{m\ne 0}$ & $\mu_r^{m=0}$ &  $\mu_r^{m\ne0}$ & Dynamo?
\\
\hline
C0 & 10 & 1& 1 &1 & yes
 \\
C1 & 1 &10 &1 &1 & no
 \\
 \hline
P0 & 1 &1 &1000 &1 & yes
 \\
P1 & 1 &1 &1 &1000 & no
 \\
\hline
\end{tabular}
\end{center}
\caption{Results of MHD simulations with different magnetic properties depending on the azimuthal order $m$ ($\hat{h}=0.1$).}
\label{tab:mag_mode}
\end{table}

To confirm this idea, we perform the following numerical experiment: 
the wall conductivity or permeability is set to different values for different azimuthal orders, \mbox{$m=0$} or \mbox{$m \ne 0$}
and the effect on dynamo action is studied. 
This is an easy modification to the code since the azimuthal direction is solved spectrally using a spherical harmonics decomposition.
The results are presented in Table~\ref{tab:mag_mode}.
In Case C0, the wall conductivity is enhanced for the mode $m=0$ but not for the modes \mbox{$m \neq 0$}
so that only the axisymmetric magnetic field sees the discontinuity of conductivity at the fluid-wall
interface. 
In this case, an axisymmetric magnetic field very similar to the dynamo case C is produced.
In Case C1, the wall conductivity enhancement is reversed and only the non-axisymmetric magnetic field sees the discontinuity of conductivity.
Then, the system fails to produce a dynamo. Similar results are obtained for mode-assigned permeability (Cases P0 and P1). 
The nonlinearity of the system implies that the amplification 
of a given magnetic mode affects the other non-amplified modes and vice versa, 
so the interpretation of the results of this numerical experiment is not entirely straightforward.
Nonetheless, the results of both Cases C0 and C1 taken together (and Cases P0 and P1 together) 
imply that this nonlinear cascade is inefficient in these cases, and therefore 
contribute further evidence to the conjecture that
the mechanism of generation of the axisymmetric poloidal field
does not require the non-axisymmetric magnetic field to be modified by the presence of the wall.

\begin{table}
\begin{center}
\begin{tabular}{ | c | c | c | c | c | c | c | c | c | c |}
\hline
Case & \multicolumn{2}{|c|}{$\sigma_r^{m=0}$} & \multicolumn{2}{|c|}{$\sigma_r^{m\ne 0}$} & 
 \multicolumn{2}{|c|}{$\mu_r^{m=0}$} &   \multicolumn{2}{|c|}{$\mu_r^{m\ne0}$} & Dynamo?
\\
\hline
 & pol & tor &pol & tor & pol & tor &pol &tor &
\\
\hline
C0P & 10 & 1& 1 &1 &1 &1 &1 & 1 & no
 \\
C0T & 1 & 10 & 1 &1 &1 &1 &1 & 1 & yes
 \\
 \hline
P0P & 1 & 1& 1 &1 &1000 &1 &1 & 1 & yes
 \\
P0T & 1 & 1& 1 &1 &1 &1000 &1 & 1 & no
\\
\hline
\end{tabular}
\end{center}
\caption{Results of MHD simulations with different magnetic properties for the poloidal and toroidal component and
depending on the azimuthal order $m$ ($\hat{h}=0.1$).}
\label{tab:mag_mode2}
\end{table}

To further narrow down the role of the wall we attempt to distinguish between its effect on $\overline{B_P}$ and $\overline{B_T}$
by performing another set of modified simulations where
we set different values of the conductivity and the permeability for the poloidal and toroidal $m=0$ modes
while the relative conductivity and permeability are kept equal to 1 for the modes $m>0$.
The results are presented in Table~\ref{tab:mag_mode2}.
A dynamo is only obtained in Cases C0T and P0P.
Again the nonlinearity of the system demands some caution in the interpretation of these results,
but these special cases taken together lend further credence to
 our theories about the role of the wall:
an enhanced conductivity amplifies $\overline{B_T}$ directly, whereas an enhanced permeability 
provokes a strengthening of $\overline{B_P}$ in the shear layer, and so
a direct enhancement of the $\omega$ effect.
No further action of the conducting wall on the other components of the magnetic field is required.

\section{Conclusions and discussion}
Through a series of high resolution simulations, we have studied dynamos driven by boundary forcing in a spherical shell geometry, 
and the effects of varying independently the thickness, electrical conductivity and magnetic permeability of the outer wall.

For an homogeneous system (same magnetic permeability and conductivity in the fluid and the wall) 
with a magnetic Prandtl number $\Pm_f=0.01$, 
the flow is unable to sustain a magnetic field at the forcing used (corresponding to $\Rey=48193$, about 200
times the critical forcing for hydrodynamical non-axisymmetric instabilities of the base flow).
For a wall thickness $h=0.1 r_o$, increasing the wall conductivity, $\sigma_w$, by a factor 10 or the wall magnetic permeability, $\mu_w$, 
by a factor 1000 creates a dynamo.
The effects of high $\sigma_w$ and high $\mu_w$ are clearly different on the dynamo threshold,
so the decrease of the magnetic diffusivity in the wall, \mbox{$\eta_w=1/(\sigma_w \mu_w)$}, is not the 
controlling parameter of this problem.
The favorable roles of large wall thickness, conductivity and magnetic permeability on dynamo action 
obtained in our numerical simulations
are in agreement with previous numerical studies using different geometry, different flows and 
in some cases, idealized boundary conditions \citep{Ava03,Mar03,Rav05,Gis08,Lag08,Gie10}.
In particular for a thin wall (thickness $h=0.01 r_o$), we found a good agreement with the results of \citet{Rob10}, where a similar setup is used
with an outer magnetic boundary condition valid in a thin-wall limit. 

In our numerical simulations,
in both large $\sigma_w$ and large $\mu_w$ cases, the dynamo generates a large-scale (axisymmetric) magnetic field.
The magnetic field is mostly an axisymmetric toroidal field and an axisymmetric dipolar poloidal component. 
The axisymmetric toroidal magnetic field, $\overline{B_T}$, is generated by an $\omega$ effect, 
corresponding to the radial shearing of the radial magnetic field in the shear boundary layer located
close to the outer wall. 
The wall plays an essential role in the amplification of $\overline{B_T}$ in the shear layer.
In the large $\sigma_w$ case, the discontinuity of conductivity allows 
strong radial gradient of the axisymmetric azimuthal magnetic field,
$\overline{B_{\phi}}$, or equivalently large latitudinal electric currents in the wall, shielding
the induction in the shear layer from the vacuum outside, thereby allowing stronger $\overline{B_{\phi}}$ in the fluid.
In the large $\mu_w$ case, the wall forces the poloidal magnetic field to be normal at the fluid-wall interface, imposing
strong radial magnetic field across the shear layer and therefore again the generation of stronger $\overline{B_{\phi}}$.
Similarly to the large $\sigma_w$ case, a thick wall provides a wide matching region to the vacuum condition thereby allowing a 
large amplitude of $\overline{B_T}$ in the shear layer.
By filtering the effects of the wall magnetic properties on the different magnetic modes (Section~\ref{sec:rolewall}),
we can reasonably conclude that the essential role of the wall on the dynamo is to allow for large 
axisymmetric toroidal field $\overline{B_T}$
in the fluid. 
The vacuum boundary condition is detrimental for the dynamo by constraining the allowable
growth of $\overline{B_T}$ and so the other magnetic components of the dynamo 
that feed from it. 
The presence of a ``shielding'' wall is therefore essential - either thick, of high conductivity or high permeability.
We conjecture that without these conditions, even at higher forcings no dynamo will be found 
if no hydrodynamical bifurcation occurs and for fixed magnetic diffusivity of the fluid.
However we emphasize that this argument applies only for shear flows where the $\omega$ effect occurs adjacent to the 
outer boundary.
 
The axisymmetric poloidal magnetic field, $\overline{B_P}$, is mostly an axial dipole, and is generated in the 
equatorial belt, where non-axisymmetric motions are strongest.
A coherent emf is produced by a narrow range of non-axisymmetric modes with azimuthal symmetry $5\leq m \leq 14$.
The velocity modes are sheared by the large latitudinal gradients of the zonal flow $\overline{u_{\phi}}$
in the equatorial region. This shearing is essential to produce appropriate azimuthal phase shifts between radial and 
latitudinal velocity components, leading to non-zero azimuthal average of the cross product between velocity
and magnetic field of same azimuthal symmetry. This emf has a significant time-average only
for a few modes which are selected by the amplitude of the non-axisymmetric
velocity, their azimuthal and latitudinal extent and the latitudinal gradients of the zonal flow.
Since all these quantities vary with the strength and the geometry of the forcing, we expect different 
non-axisymmetric dynamo modes to be selected for different forcings.

In the more general context of dynamo theory, we note that this numerical dynamo model operates similarly to 
the asymptotic dynamo model of \citet{Bra64} where a large zonal flow generates
a strong axisymmetric toroidal magnetic field, and a small deviation of the flow from axisymmetry
may be sufficient to produce an axisymmetric poloidal magnetic field to overcome
Cowling's theorem.

Having established how our dynamo model operates,  
it is important to discuss
how our results relate to the observations of the VKS experiment.
In this work, we use a spherical geometry for numerical convenience. In this geometry and at the large
forcings studied here, the shear exerted by the zonal flow is located in a
viscous boundary layer at the outer wall, 
yielding significant influence of the wall parameters on the dynamo action.  
In the cylindrical von K\'arm\'an setup used in the VKS experiment and at Reynolds number of the order of $10^6$, 
velocity measurements in water show evidence that the largest axial gradients of the shear layer are located in the 
equatorial mid-plane between the two counter-rotating disks \citep{Mar03}. 
Consequently, an ``axial'' $\omega$ effect, the shearing of the axial magnetic field lines, is thought to 
be operating in the equatorial mid-plane of the cylinder \citep{Bou02}.
This is a major difference between our numerical work and the VKS experiment. If the shear layer is far from
the outer boundary, then the amplification by the wall of the toroidal field 
may not be operating in the VKS experiment.
In the VKS experiment, the generation of the axisymmetric poloidal magnetic field is usually described as 
the result of an $\alpha$ effect produced by the helical vortices present between the blades fixed on the
rotating flat disks \citep{Pet07}. In numerical studies,
this effect has been parametrized either by adding a source term in the magnetic induction equation of the 
mean field \citep{Lag08,Gie10} or by using an analytical formulation
of non-axisymmetric flow \citep{Gis09}.
In this scenario, $\overline{B_P}$ is produced 
close to the disks, and the generation mechanism is thought to
be helped by the high magnetic permeability or electrical conductivity of the disks. 
In our work, somewhat differently, we found that $\overline{B_P}$ is produced in the equatorial belt by fluctuating non-axisymmetric motions, 
and that the wall parameters do not affect this mechanism.

The discrepancy in the location of the active dynamo regions, and therefore the potentially different
role of the wall on the different steps of the dynamo feedback loop, 
may reveal the importance of the geometry of the container or the presence
of blades on the rotating walls in these dynamo experiments, both physical and numerical. 
However the different approaches used to calculate the induction effects 
(be they self-consistent or parametrized) in the various numerical codes used to simulate the physical experiment
may also explain part of the discrepancy.
In a forthcoming work, we will study the consequences of the geometry of the forcing on dynamo action,
still using a spherical geometry but varying the latitudinal profile of angular velocity of the wall. 
Here, we find that the latitudinal gradients of the zonal flow play an important role in the selection of the
non-axisymmetric dynamo modes sustaining the axial dipole. 
Forcing confined closer to the poles may simulate the experimental forcing more realistically.
A further interesting question is whether there 
are certain configurations of the forcing which cannot 
produce dynamo action.

\section*{Acknowledgments}
The authors would like to thank P.~Cardin, P.~H.~Diamond, P.~Garaud, G.~A.~Glatzmaier, D.~W.~Hughes, S.~M.~Tobias, G.~R.~Tynan, K.~White, T.~S.~Wood 
for useful discussions
and two anonymous referees for improving the manuscript.
Financial support was provided by the Center for Momentum Transport and Flow Organization (CMTFO), a Plasma Science Center
sponsored by the US Department of Energy (DoE) Office of Fusion Energy Sciences and the American Recovery and Reinvestment Act (ARRA) 2009.
This research was further supported by an allocation of advanced computing resources provided by the 
National Science Foundation (NSF).  The computations were performed on the NSF Teragrid/XSEDE machine Kraken at the National Institute for Computational Sciences (NICS)
and on University of California Santa Cruz (UCSC) supercomputer Pleiades purchased under NSF MRI grant AST-0521566.

\appendix

\section{Implementation of the magnetic boundary conditions}
\label{app:BC}

In this appendix, we derive the magnetic boundary conditions at an interface located 
at the radius $r_o$ between the fluid and the wall (denoted by the subscript $f$ and $w$ respectively)
of different magnetic permeability, $\mu$, or electrical conductivity, $\sigma$. 
The magnetic properties in each media are constant.
The relative permeability and conductivity are $\mu_r=\mu_w/\mu_0$ and $\sigma_r=\sigma_w/\sigma_f$.

The magnetic field is decomposed into poloidal and toroidal vectors:
\begin{equation}
  \vect{B} = \vect{\nabla} \times \vect{\nabla} \times (B_P \vect{r}) + \vect{\nabla} \times (B_T \vect{r}) ,
\end{equation}
where $B_P$ and $B_T$ are the poloidal and toroidal scalars.
The spherical components of $\vect{B}$ can be expressed in function of these two scalars by
\begin{eqnarray} 
B_r &=& \frac{1}{r} L_2(B_P) ,
\\
B_{\theta} &=& \frac{\partial}{\partial \theta} \left[ \frac{1}{r} \frac{\partial}{\partial r} \pleft r B_P \pright \right] 
                 + \frac{1}{\sin \theta} \frac{\partial B_T}{\partial \phi} ,
\\ 
B_{\phi} &=& \frac{1}{\sin \theta} \frac{\partial}{\partial \phi} \left[ \frac{1}{r} \frac{\partial}{\partial r} \pleft r B_P \pright \right] 
                - \frac{\partial B_T}{\partial \theta} ,
\end{eqnarray}
with $L_2$ the angular laplacian operator,
\begin{eqnarray}
 L_2 = - \frac{1}{\sin \theta} \frac{\partial}{\partial \theta} \pleft \sin \theta \frac{\partial}{\partial \theta} \pright
	- \frac{1}{\sin^2 \theta} \frac{\partial^2}{\partial \phi^2} .
\end{eqnarray}

Through a magnetic permeability discontinuity, the continuity of the radial magnetic field (Equation~\ref{eq:Bnormal}), 
and the discontinuity of the tangential magnetic field (Equation~\ref{eq:Btan}) yields
\begin{eqnarray}
	\left. B_P \right|_w &=& \left. B_P \right|_f ,
	\label{eq:cont_Bp}
	\\
	\left. B_T \right|_w &=& \mu_r \left. B_T \right|_f ,
	\\
	\left. \frac{\partial r B_P}{\partial r} \right|_w &=& \mu_r \left. \frac{\partial r B_P}{\partial r} \right|_f .
	\label{eq:cont_Bs}
\end{eqnarray}

The electric current density, $\vect{j}$, is divergence-free and can also be decomposed into poloidal and 
toroidal scalars, $j_P$ and $j_T$.
Through an electrical conductivity discontinuity, the continuity of the radial electric current density (Equation~\ref{eq:jnormal}), 
and the discontinuity of the tangential electric current density (Equation~\ref{eq:jtan}) yields
\begin{eqnarray}
	\left. j_P \right|_w &=& \left. j_P \right|_f ,
	\\
	\left. j_T \right|_w &=& \sigma_r \left. j_T \right|_f ,
	\\
	\left. \frac{\partial r j_P}{\partial r} \right|_w &=& \sigma_r \left. \frac{\partial r j_P}{\partial r} \right|_f .
	\label{eq:cont_djpdr}
\end{eqnarray}
$j_P$ and $j_T$ are related to the $B_P$ and $B_T$ by
\begin{eqnarray}
	j_P &=&  \frac{B_T}{\mu},
	\label{eq:jp}
	\\
	j_T &=& \frac{1}{r^2} L_2 \frac{B_P}{\mu} - \frac{1}{r} \frac{\partial}{\partial r} \frac{r B_S}{\mu},
	\label{eq:jt}
	\\
	j_S &=& \frac{1}{r} \frac{\partial}{\partial r} \frac{r B_T}{\mu},
	\label{eq:js}
\end{eqnarray}
where we use the spheroidal scalar for a divergence-free vector, 
\begin{eqnarray}
	j_S = \frac{1}{r} \frac{\partial r j_P}{\partial r} .
\end{eqnarray}
To solve the magnetic induction equation, we need to evaluate second order radial derivatives on our irregular radial grid.
To obtain the finite difference scheme of second order in the vicinity of $r_o$, a radial function $g$ is expanded using 
Taylor's formula:
\begin{eqnarray}
	g(r_o-dr^-) &=& g(r_o^-) - g'(r_o^-) dr^- + g''(r_o^-) \frac{(dr^-)^2}{2} + \mathcal{O}((dr^-)^3) ,
	\label{eq:Taylor_m}
	\\
	g(r_o+dr^+) &=& g(r_o^+) + g'(r_o^+) dr^+ + g''(r_o^+) \frac{(dr^+)^2}{2} + \mathcal{O}((dr^+)^3),
\end{eqnarray}
where $r_o^-$ ($r_o^+$) is located infinitely close to $r_o$ on the fluid side (wall side respectively) and 
$dr^-$ ($dr^+$) is the radial incremental step between $r_o$ and his neighbor on the radial grid on the side $f$ ($w$ respectively).

The function $g$ is chosen such as it is continuous at the interface, while its radial derivatives are not:
\begin{eqnarray}
	g(r_o^+) & = & g(r_o^-),
	\\
	g'(r_o^+) & = & \gamma g'(r_o^-),
	\\
	g''(r_o^+) & =& \alpha g''(r_o^-) + \beta g(r_o^-) + q,
	\label{eq:g2_cont}
\end{eqnarray}
where $q$ represents the possible contribution from the non-linear terms.

The second-order derivative of $g$ is obtained by taking the combination $g(r_o - dr^-)\gamma dr^+ + g(r_o+dr^+) dr^-$,
then applying the identities~(\ref{eq:Taylor_m}) -- (\ref{eq:g2_cont}) and rearranging, which leads to
\begin{eqnarray}
	g''(r_o^-) &=& \frac{2}{dr^+ dr^- (\alpha dr^+ + \gamma dr^-)} \left[ dr^- g(r_o +dr^+) + \gamma dr^+ g(r_o -dr^-) \right. 
	\nonumber
	\\
			&& \left. - (dr^- + \gamma dr^+ ) g(r_o^-) -  \frac{dr^- (dr^+)^2}{2} \beta  g(r_o^-) \right] -\frac{dr^+}{\gamma dr^- + \alpha dr^+} q.
\label{eq:g2}
\end{eqnarray}

\subsection{Poloidal component}
\label{secapp:_pol_linear}
The evolution equation for the poloidal scalar, $B_P$, is obtained by taking the dot product of the magnetic induction equation with $\vect{r}$:
\begin{eqnarray}
  \pdt{B_P} = - \frac{1}{\sigma} \left[
				\frac{1}{r^2}L_2 \frac{B_P}{\mu} 
				- \frac{1}{r}\frac{\partial}{\partial r} \pleft 
					\frac{1}{\mu} \frac{\partial}{\partial r} (r B_P) \pright
				\right]
			+ f_P ,
\label{eq:dBpdt}
\end{eqnarray}
where $f_P$ contains the non-linear terms.
We use a spherical harmonics expansion for $B_P$,
\begin{eqnarray}
	B_P (r,\theta,\phi) = \sum\limits_{l=0}^{\infty} \sum\limits_{m=-l}^{l} p_l^m(r) Y_l^m(\theta,\phi) ,
\end{eqnarray}
to solve the angular laplacian operator:
\begin{eqnarray}
	L_2 B_P = \sum\limits_{l} \sum\limits_{m} l(l+1) p_l^m Y_l^m,
\end{eqnarray}
where $Y_l^m$ is the spherical harmonic function of degree $l$ and order $m$.

The non-linear terms are:
\begin{eqnarray}
	f_P &=& \frac{1}{l(l+1)} \vect{r} \cdot \nabla \times (\vect{u} \times \vect{B} ) , \\
	&=& \frac{1}{l(l+1) \sin \theta} \left( \frac{\partial}{\partial \theta} \sin \theta \left[ \vect{u} \times \vect{B} \right]_{\phi}
							- \frac{\partial}{\partial \phi} \left[ \vect{u} \times \vect{B} \right]_{\theta} \right).
\end{eqnarray}
The cross-product \mbox{$\vect{u} \times \vect{B}$} is calculated in spatial space using 
\mbox{$\vect{u}=(u_r,u_{\theta},u_{\phi})$} and $\vect{B}=(B_r,B_{\theta},B_{\phi})$.
At the fluid-wall interface, the velocity in the fluid matches continuously to the no-slip boundary condition, 
\mbox{$\vect{u}(r_o)=(0,0,\overline{u_{\phi}}(r_o))$}
with the azimuthal velocity in the wall \mbox{$\overline{u_{\phi}} = r \sin \theta \cos \theta$}.
The spherical components of the cross-product in the wall are 
\begin{eqnarray}
	\left[ \vect{u} \times \vect{B} \right]_r &=& -\overline{u}_{\phi} B_{\theta} ,
	\\
	\left[ \vect{u} \times \vect{B} \right]_{\theta} &=& \overline{u}_{\phi} B_r ,
	\\ 
	\left[ \vect{u} \times \vect{B} \right]_{\phi} &=& 0 . 
\end{eqnarray}
The continuity of the velocity and $B_r$ across the interface therefore insures the continuity of the non-linear term $f_P$.

The difficulty in the numerical implementation of Equation~(\ref{eq:dBpdt}) is therefore to calculate the second-order radial derivative
of $r p_l^m$. Using $g=r p_l^m$ and Equation~(\ref{eq:cont_Bs}), 
\begin{eqnarray}
	g'(r_o^+) = \mu_r g'(r_o^-).
\label{eq:g1_Bp}
\end{eqnarray}
The right-hand side of Equation~(\ref{eq:dBpdt}) must be continuous across the interface yielding,
\begin{eqnarray}
	g''(r_o^+) = \sigma_r \mu_r g''(r_o^-) + g(r_o^-) \frac{l(l+1)}{r^2_o} (1 -\sigma_r \mu_r).
\label{eq:g2_Bp}
\end{eqnarray}
We can now use the identity~(\ref{eq:g2}) to calculate $g''(r_o^-)$ from the values of $g$ at $r=r_o-dr^-$, 
$r=r_o$ and $r=r_o+dr^+$. This allow us to evaluate the right-hand side of
the linear evolution equation for $p_l^m$ at the point $r_o^-$.
The radial scheme in the wall uses the point at $r_o^+$ with $p_l^m(r_o^+)=p_l^m(r_o^-)$.

\subsection{Toroidal component}
The evolution equation for the toroidal scalar, $B_T$, is obtained by 
taking the dot product of the curl of the magnetic induction equation with $\vect{r}$:
\begin{eqnarray}
  \pdt{B_T} = - \frac{1}{r^2} L_2 \frac{1}{\sigma} \frac{B_T}{\mu}
		    + \frac{1}{r} \frac{\partial}{\partial r} 
			\frac{1}{\sigma} \frac{\partial}{\partial r}\pleft r \frac{B_T}{\mu} \pright 
			+ f_T.
\label{eq:dBtdt}
\end{eqnarray}
Again a spherical harmonics expansion is used,
\begin{eqnarray}
	B_T (r,\theta,\phi) = \sum\limits_{l=0}^{\infty} \sum\limits_{m=-l}^{l} t_l^m(r) Y_l^m(\theta,\phi) .
\end{eqnarray}

The non-linear terms are:
\begin{eqnarray}
	f_T &=& \frac{1}{l(l+1)} \vect{r} \cdot \nabla \times \nabla \times (\vect{u} \times \vect{B} ),  \\
	&=& \frac{1}{r} [\vect{u} \times \vect{B}]_r
	+ \frac{1}{l(l+1) \sin \theta}
	 \left[ \frac{\partial}{\partial \theta} \sin \theta \frac{1}{r} \frac{\partial}{\partial r} \left( r \left[ \vect{u} \times \vect{B} \right]_{\theta} \right)
		+ \frac{\partial}{\partial \phi}  \frac{1}{r} \frac{\partial}{\partial r} \left( r \left[ \vect{u} \times \vect{B} \right]_{\phi} \right) \right].
		\label{eq:fT}
\end{eqnarray}
At the interface $B_T(r_o^+)=\mu_r B_T(r_o^-)$, so the non-linear terms are continuous if $f_T(r_o^+) = \mu_r f_T(r_o^-)$. 
The first term of right-hand side of~(\ref{eq:fT}), $[\vect{u} \times \vect{B}]_r=  -\overline{u}_{\phi} B_{\theta}$, 
respects this continuity condition.
However the second term on the right-hand side of~(\ref{eq:fT}) does not respect this condition because of
the discontinuity of the radial gradient of the velocity across the interface. The discontinuity of $f_T/\mu$
must therefore be taken into account in the numerical implementation of Equation~\ref{eq:dBtdt}.

We use $g=r t_l^m / \mu$. Using $j_P = B_T/\mu$ and Equation~(\ref{eq:cont_djpdr}), we obtain
\begin{eqnarray}
	g'(r_o^+) = \sigma_r g'(r_o^-) .
\label{eq:g1_Bt}
\end{eqnarray}
The right-hand side of Equation~(\ref{eq:dBtdt}) divided by $\mu$ must be continuous across the interface yielding,
\begin{eqnarray}
	g''(r_o^+) = \sigma_r \mu_r g''(r_o^-) +g(r_o^-) \frac{l(l+1)}{r^2_o} (1-\sigma_r \mu_r) 
	+ \sigma_w r_o \left( \mu_r f_T(r_o^-) - f_T(r_o^+) \right).
\label{eq:g2_Bt}
\end{eqnarray}
Using the identity~(\ref{eq:g2}) to calculate $g''(r_o^-)$ from the values of $g$ at $r=r_o-dr^-$, 
$r=r_o$ and $r=r_o+dr^+$, we can evaluate the evolution equation for $t_l^m$ at the point $r_o^-$, 
\begin{eqnarray}
	\pdt{t_l^m (r_o^-)} = - \frac{1}{\mu_f \sigma_f}  \frac{l(l+1)}{r_o^2} t_l^m(r_o^-) 
					 + \frac{1}{r_o \sigma_f} \tilde{g}''(r_o^-)
					 + \frac{dr^+ f_T(r_o^+) + dr^-f_T(r_o^-)}{dr^- + \mu_r dr^+},
\label{eq:tlm_final}
\end{eqnarray}
where
\begin{eqnarray}
	\tilde{g}''(r_o^-) &=& \frac{2}{dr^+ dr^- \sigma_r(dr^- + \mu_r dr^+)} \left[ \sigma_r dr^+ g(r_o-dr^-) +dr^- g(r_o+dr^+) \right.
				\\ && \left. - g(r_o^-) \left(\sigma_r dr^+ + dr^- + \frac{dr^{+2} dr^-}{2} \frac{l(l+1)}{r_o^2} (1-\sigma_r \mu_r) \right) \right] .
\end{eqnarray}
The contribution of the non-linear terms to the evolution of $t_l^m $ at $r=r_o^-$ 
is therefore evaluated through a weighted average of $f_T(r_o^-)$ and $f_T(r_o^+)$ 
(third term on the right-hand side of Equation~\ref{eq:tlm_final}),
which insures continuity of $B_T/\mu$ and of $(\sigma^{-1})\partial_r (r B_T\mu^{-1})$.

The radial scheme in the wall uses the point at $r_o^+$ with $t_l^m(r_o^+)=\mu_r t_l^m(r_o^-)$.

\subsection{Match to external potential field}
The vacuum boundary condition at $r_v=r_o+h$ corresponds to $\vect{j} = 0$. 
The interface between the wall ($w$) and the vacuum ($v$) corresponds to a jump of magnetic permeability
and electrical conductivity. 
The toroidal scalar vanishes in the vacuum, so the boundary condition at $r_v^-$, the point infinitely close 
to the interface on the side of the wall is simply,
\begin{eqnarray}
	t_l^m (r_v^-) = 0 .
\end{eqnarray}
In the vacuum, the poloidal scalar follows
\begin{eqnarray}
	\frac{\partial p_l^m}{\partial r} + \frac{l+1}{r} p_l^m = 0 .
\end{eqnarray}
Using Equations~(\ref{eq:cont_Bp}) and~(\ref{eq:cont_Bs}), the boundary condition for the poloidal scalar is
\begin{eqnarray}
	\left. \frac{\partial p_l^m}{\partial r} \right|_{r_v^-} = -\frac{\mu_r l+1}{r} p_l^m(r_v^-).
\end{eqnarray}


%

\end{document}